\documentclass[twocolumn,
  nobibnotes,
  aps,
  pra,
  showpacs,
  amsmath,
  amssymb,
  superscriptaddress,
  floatfix,
  10pt]{revtex4-1}

\usepackage{graphicx}
\usepackage{bm}
\usepackage{amsmath}

\usepackage{color}

\usepackage[T1]{fontenc}
\usepackage{bbold}

\begin{document}

\title{Transversal magnetoresistance and Shubnikov-de Haas oscillations in Weyl semimetals}

\author{J. Klier}

\affiliation{Institut f\"ur Nanotechnologie, Karlsruhe Institute of Technology, 76021 Karlsruhe, Germany}

\affiliation{\mbox{Institut f\"ur Theorie der Kondensierten Materie, Karlsruhe Institute of Technology, 76128 Karlsruhe, Germany}}

\author{I.V. Gornyi}

\affiliation{Institut f\"ur Nanotechnologie, Karlsruhe Institute of Technology, 76021 Karlsruhe, Germany}

\affiliation{\mbox{Institut f\"ur Theorie der Kondensierten Materie, Karlsruhe Institute of Technology, 76128 Karlsruhe, Germany}}

\affiliation{A. F. Ioffe Physico-Technical Institute, 194021 St.~Petersburg, Russia}

\author{A.D. Mirlin}

\affiliation{Institut f\"ur Nanotechnologie, Karlsruhe Institute of Technology, 76021 Karlsruhe, Germany}

\affiliation{\mbox{Institut f\"ur Theorie der Kondensierten Materie, Karlsruhe Institute of Technology, 76128 Karlsruhe, Germany}}

\affiliation{Petersburg Nuclear Physics Institute, 188350 St.~Petersburg, Russia}

\date{\today}%
\begin{abstract}
We explore theoretically the magnetoresistance of Weyl semimetals in transversal magnetic fields away from charge neutrality. The analysis
within the self-consistent Born approximation is done for the two different models of disorder: (i) short-range impurties and (ii) charged (Coulomb) impurities. For these models of disorder, we calculate the conductivity away from charge neutrality point as well as the Hall conductivity,
and analyze the transversal magnetoresistance (TMR) and Shubnikov-de Haas oscillations for both types of disorder.
We further consider a model with Weyl nodes shifted in energy with respect to each other (as found in various materials)
with the chemical potential corresponding to the total charge neutrality.
In the experimentally most relevant case of Coulomb impurities, we find in this model a large TMR in a broad range of quantizing magnetic fields.
More specifically, in the ultra-quantum limit, where only the zeroth Landau level is effective,
the TMR is linear in magnetic field. In the regime of moderate (but still quantizing) magnetic fields, where the higher Landau levels are relevant,
the rapidly growing TMR is supplemented by strong Shubnikov-de Haas oscillations, consistent with experimental observations.
\end{abstract}
\maketitle

\section{Introduction}

One of the central research directions in condensed matter physics addresses topological materials and structures.
Recently, a novel type of topological materials has received much attention: Weyl and Dirac semimetals.
The quasiparticle spectrum near the nodal point of a Dirac semimetal is described by a three-dimensional (3D) $4\times4$ Dirac Hamiltonian
where excitations close the crossing point of valence and conduction bands disperse linearly.
The materials Cd$_{3}$As$_{2}$ \cite{RIS_0} and Na$_{3}$Bi \cite{Liu21022014} represent experimental realizations of Dirac semimetals.
For either broken spatial inversion or time-reversal symmetry, the four-component solution of the Dirac equation splits into two independent
two-component Weyl fermions of opposite chirality with the Weyl points in the spectrum located at distinct momenta.
Recent experiments classify TaAs~\cite{2015arXiv150204684L,2015arXiv150203807X}, NbAs~\cite{2015arXiv150401350X}, TaP~\cite{Xue1501092},
and NbP~\cite{RIS_5} as Weyl semimetals. Further promising candidates for Weyl semimetals include pyrochlore iridates~\cite{PhysRevB.83.205101} and topological insulator heterostructures~\cite{PhysRevLett.107.127205}.
In the rest of the paper, we will use the term ``Weyl semimetal'' in a broader sense, including also the degenerate case of Dirac semimetals.

Transport properties of Weyl semimetals are highly peculiar. For recent theoretical studies, see,  e.g., Refs.~\cite{PhysRevB.84.235126,PhysRevLett.108.046602,reviewVish,0953-8984-27-11-113201,PhysRevB.89.054202,
PhysRevLett.113.026602,PhysRevB.88.104412,PhysRevB.89.014205, PhysRevB.91.035133, PhysRevLett.114.257201,PhysRevLett.113.247203, PhysRevB.91.245157, PhysRevLett.114.166601, 2015arXiv150507374S,PhysRevB.91.035202,PhysRevB.91.195107,PhysRevB.83.205101,2016arXiv160801286Z} and references therein.
An important aspect of the transport properties is the appearance of a disordered critical point within the perturbative analysis. Below the disordered critical point (i.e., for sufficiently weak disorder), the density of states vanishes quadratically in energy around the Weyl point within the perturbation theory. Non-perturbative treatment yields an exponentially small density of states at the Weyl point.
In the strong disorder regime, the density of states is finite at the Weyl point already without invoking exponentially small contributions.

The transport in Weyl semimetals reveals a particularly interesting and rich physics when an
external magnetic field is applied.
One reason for this is the unconventional Landau quantization of Dirac fermions.
Further, a single species of Weyl fermions displays the chiral anomaly that gives rise to a possibility of controlling the valley
polarization. A strong anomalous Hall effect~\cite{reviewVish,0953-8984-27-11-113201,PhysRevLett.111.027201} and the
longitudinal magnetoresistivity~\cite{PhysRevB.88.104412, PhysRevB.89.085126, PhysRevLett.113.247203, Lucas23082016,PhysRevB.91.245157, 2015arXiv150302069G, 0953-8984-27-15-152201, 2015arXiv150606577S, RIS_5,2017arXiv170401038B} in Weyl semimetals have been predicted to originate from the chiral anomaly. Furthermore, thermoelectrical effects \cite{Gooth} and induced superconductivity \cite{Bachmanne1602983} have been studied recently, both  theoretically and experimentally.

In this paper, we present a theory of the transversal magnetoresistivity  in a Weyl semimetal away from charge neutrality point.
(The term ``transversal'' here means that the magnetic field is perpendicular to the electric field: the relevant resistivity component is $\rho_{xx}$, while the magnetic field is along the $z$ axis.)
The work is motivated by the spectacular experimental observation of a large, approximately linear transversal magnetoresistance (TMR)
in Dirac and Weyl semimetals~\cite{RIS_1, PhysRevB.92.081306, RIS_5, 2016arXiv161001413C,PhysRevB.91.041203}.
Theoretically, a linear TMR of a system with Dirac dispersion in the ultra-quantum limit
(where only the zeroth Landau level is effective) was obtained
by Abrikosov in a seminal paper, Ref.~\cite{PhysRevB.58.2788}. The crucial ingredient of this result is
the dependence of the screening of Coulomb impurities on magnetic field. In a previous work, Ref.~\cite{PhysRevB.92.205113}, we have carried out a systematic analysis of the magnetoresistivity of a Weyl semimetal at the neutrality point and for different types of disorder. Our results for the case of Coulomb impurities and in strongest magnetic fields yield the linear TMR, in agreement with  Ref.~\cite{PhysRevB.58.2788}. This is not sufficient, however, to explain experimental data since experiments are performed at non-zero electron density. A clear experimental evidence of finite density is provided by Shubnikov-de Haas oscillations (SdHO) superimposed on the background of strong linear TMR in an intermediate range of magnetic fields. It is thus a challenge to understand whether the strong quantum linear TMR and the SdHO may emerge from the theory of disordered Weyl fermions. More generally, our goal is to develop the theory of quantum magnetotransport for systems with Dirac spectrum at non-zero density (chemical potential) of carriers.

Below, we calculate the TMR and the Hall conductivity for arbitrary magnetic field $H$ and arbitrary particle density. 
Depending on their values, the dominant contribution to the TMR comes from the zeroth Landau level (LL), separated LLs, or overlapping LLs. This
includes also regimes where the SdHO can be observed. Our analysis has a certain overlap with a recent preprint, Ref.~\cite{2016arXiv160704943X}, where the Born approximation (without self-consistency) was used.
We go beyond that work by employing the self-consistent Born approximation (SCBA),
analyzing the scaling of conductivities and of TMR in various regimes, 
and discussing two models of disorder---(i) short-range impurities and (ii) charged (Coulomb) impurities.
Further, we study the TMR for two cases---fixed particle density and fixed chemical potential---and find that the results are essentially different.

In the experimentally most relevant case of Coulomb impurities and a fixed particle density, we find a large, linear TMR in the 
ultra-quantum limit, where only the zeroth Landau level is effective.
We show, that even though the analytical result for the resistivity is modified in comparison to that of Ref.~\cite{PhysRevB.58.2788} 
due to a non-zero value of the Hall conductivity, the linear-in-$H$ scaling of TMR remains valid.
In the regime of moderate (but still quantizing) magnetic fields,
where the higher Landau levels are relevant, the TMR curves contain Shubnikov-de Haas peaks whose 
amplitude grows as a power-law function ($H^{4/3}$) of magnetic field. At the same time, the ``background'' TMR (the envelope of the minima)
in such magnetic fields is negligible within the SCBA. Thus, the model with a single type of Weyl nodes
does not contain a regime where a strong TMR is supplemented by SdHO, in agreement with the numerical findings of Ref.~\cite{2016arXiv160704943X}.

\begin{figure}
\includegraphics[width=\linewidth]{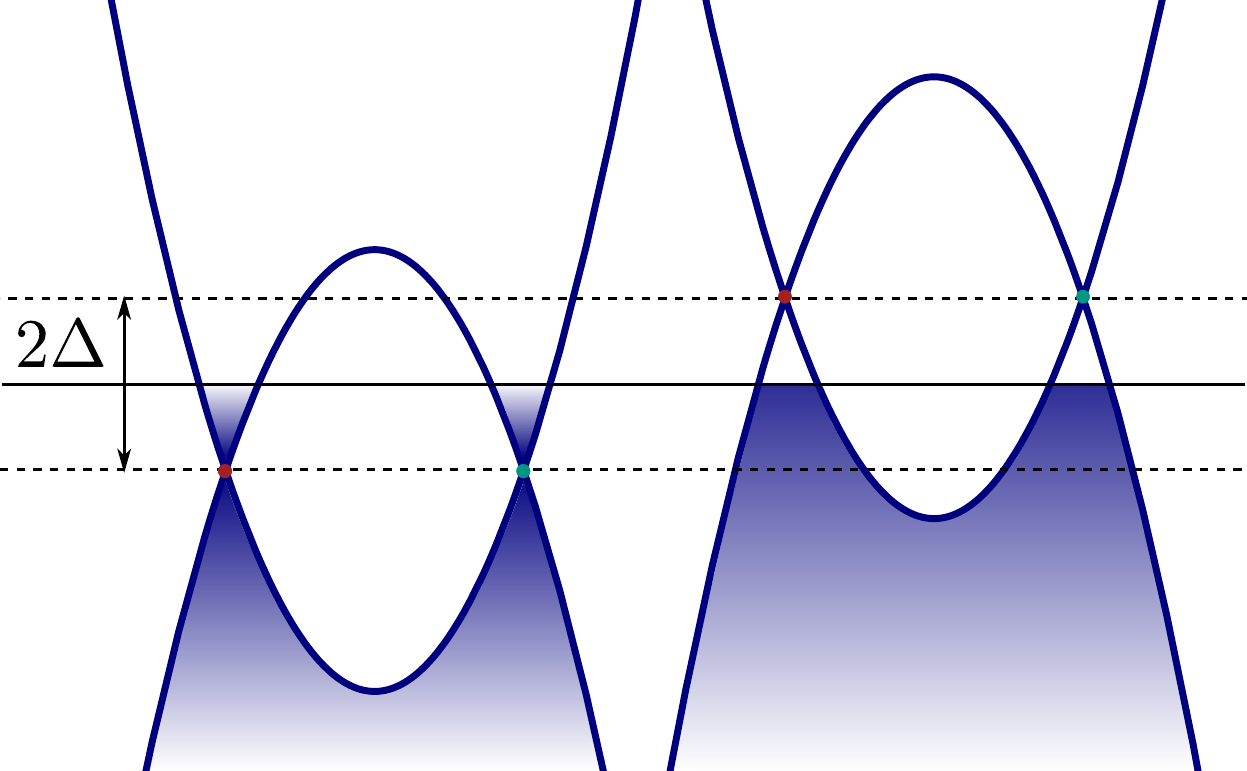}
\caption{Schematic energy band structure of the material with two pairs of Weyl nodes
shifted in energy with respect to each other.
The carriers belonging to the two pairs of nodes have a chemical potential (counted from the corresponding node) 
of $\Delta$ (electron-type carriers) and $-\Delta$ (hole-type), respectively.
Therefore, the system is at the total charge compensation point.}
\label{energybands}
\end{figure}

We further consider a model with Weyl nodes shifted in energy with respect to each other, 
with the chemical potential corresponding to the total charge neutrality, as illustrated in Fig.~\ref{energybands}.
Such type of spectrum has been found in various materials both experimentally and by first-principle calculations, 
see, e.g., Refs. \cite{RIS_5, 2016arXiv161001413C}.
In this situation, the total Hall conductivity is zero, whereas the shifted pairs of Weyl nodes are characterized by
equal carrier (electrons and holes, respectively) densities.  
For Coulomb impurities, we find in this model a large TMR in a broad range of quantizing magnetic fields.
In the ultra-quantum limit, where only the zeroth Landau level is effective,
the TMR is again linear in magnetic field. At lower magnetic fields, in the regime of separated LLs, strong
SdHO are superimposed on top of a rapidly growing background TMR, in contrast to the case
of non-shifted Weyl nodes. Specifically, the envelope of the minima of TMR behaves as $H^{2/3}$, 
while the maxima evolve as $H^2$. The overall behavior of the TMR resembles that found in experiments:
with increasing magnetic field the (almost linear) TMR shows SdHO and crosses over into a purely linear TMR
with no SdHO. Such a behavior emerges when the conductivity $\sigma_{xx}$ in a strong magnetic field is
larger (due to the compensation between the shifted nodes) than the total Hall conductivity $\sigma_{xy}$. 
This can be realized for shifted Weyl nodes away from the charge neutrality point (where the Hall resistivity is finite), provided that the
concentrations of positively and negatively charged impurities are close to each other.
  
The analysis in this paper is performed in the framework of the SCBA for non-interacting fermions. 
This discards other possible contributions to the TMR, including the classical memory effects 
(as discussed in the context of Weyl semimetals in a recent paper, Ref.~\cite{PhysRevB.92.180204}) 
and interaction-related mechanisms. We will return to a discussion of such magnetoresistance mechanisms 
in the end of the paper.

The paper is organized as follows.
Section \ref{sec:model} is devoted to an introduction to the model of impurity scattering.
In Sec.~\ref{sec:Sxx}, we calculate the conductivity $\sigma_{xx}$ away from charge neutrality in a finite
transverse magnetic field for the model of white-noise disorder.
Section~\ref{sec:Hall} presents the analysis of the Hall conductivity for the clean case and for the white-noise disorder.
In Sec.~\ref{sec:magneto}, we use the obtained results to calculate and analyze the TMR.
In Sec.~\ref{sec:charged}, we extend our analysis to the case of charged impurities.
Section \ref{sec:shifted} discusses the TMR at the total charge compensation point for the pairs of Weyl nodes shifted in energy
with respect to each other.
We summarize our findings and discuss the experimental relations to experiments in Sec.~\ref{sec:summary}.
Throughout the paper we set $\hbar=c=k_B=1$.

\section{Model}
\label{sec:model}

In this section, we introduce the framework~\cite{PhysRevB.92.205113} for studying disordered Weyl fermions that will be used throughout the paper.
We start from the Hamiltonian for a single Weyl fermion in the presence of a finite magnetic field directed along the $z$ axis.
The Hamiltonian in the Landau gauge for a clean system is given by
\begin{equation}
\mathcal{H}\left(\textbf{p}\right)=\int d^{3}\textbf{r}\,\Psi^{\dagger}(\textbf{r})\, v\bm{\sigma}\left(\textbf{p}-\frac{e}{c}\textbf{A}\right)\Psi(\textbf{r}),
\end{equation}
where $\textbf{p}$ is the momentum operator, $v$ is the velocity,
$\boldsymbol{\sigma}$ denotes the Pauli matrices and  $\textbf{A}(\textbf{r})=(0, Hx, 0)$ is the vector potential.

Now we include disorder. The impurity scattering generates a self-energy $\hat{\Sigma}(\textbf{p}, \varepsilon)$ in the (impurity-averaged) Green's function, which reads
\begin{equation}\label{green}
\hat{G}(\textbf{p}, \varepsilon)=\left\langle\frac{1}{\varepsilon-\mathcal{H}}\right\rangle
=\frac{1}{\varepsilon-v\boldsymbol{\sigma}\cdot\left(\textbf{p}-\frac{e}{c}\textbf{A}\right)-{\hat{\Sigma}}(\textbf{p}, \varepsilon)}.
\end{equation}
The Green's function is a matrix in the pseudospin space (in which the Pauli matrices $\sigma$ operate).
We will assume that the disorder potential is diagonal in both spin and pseudospin indices and neglect scattering between different Weyl nodes.
Clearly, in the absence of internode scattering, the structure in the node space will be trivial for all quantities;
the  density of states and the conductivities calculated below are those per Weyl node.
Under these assumptions, the pointlike impurity potential has the form
\begin{equation}
\hat{V}_{\text{dis}}(\textbf{r})=u_{0}\sum_{i}\delta(\textbf{r}-\textbf{r}_{i})
\mathbb{1},
\end{equation}
where $\mathbb{1}$ is the unit matrix in the pseudospin space.

In view of the matrix structure of the impurity potential $\hat{V}_{\text{dis}}(\textbf{r})$,
the impurity correlator $\hat{W}$ becomes a rank-four tensor.
Within the self-consistent Born approximation (SCBA), the self-energy reads
\begin{equation}\label{scba}
\Sigma_{\alpha\beta}(\textbf{r},\textbf{r}')=
\int\frac{d^{3}q}{(2\pi)^{3}}W_{\alpha\gamma\beta\delta}(\textbf{q})
e^{i\textbf{q}\cdot(\textbf{r}-\textbf{r}')}G_{\gamma\delta}(\textbf{r},\textbf{r}').
\end{equation}
For a diagonal impurity potential, the impurity correlator is diagonal as well, which is expressed as
\begin{equation}\label{correlator}
W_{\alpha\gamma\beta\delta}(\textbf{q})=\gamma
\delta_{\alpha\gamma}\delta_{\beta\delta},
\end{equation}
where $\gamma=n_\text{imp}u_0^2$.
We will later generalize the results obtained for white-noise disorder (\ref{correlator})
to the case of Coulomb impurities.
Similarly to the case of zero magnetic field, we introduce a parameter $\beta$ defined as
\begin{align}
\beta=\frac{\gamma\Lambda}{2\pi^2v^2},
\end{align}
where $\Lambda$ is the ultraviolet energy cutoff for energy (band width).
In the following, we will mainly focus on the case of not too strong disorder, $\beta<1$.

The self-energy is diagonal in the energy-band space.
However, in the presence of magnetic field, the self-energy is no longer proportional to the unit matrix.
This asymmetry originates from the asymmetry of states in the zeroth LL.
In the clean case, the states of the zeroth LL are only present in one energy band.
Note that a strong impurity scattering eliminates this asymmetry.
In what follows, it is convenient to switch to the LL representation such that
$\hat{G}=\hat{G}(\varepsilon,p_z,n)$ and $\hat{\Sigma}=\hat{\Sigma}(\varepsilon,p_z,n)$.
The diagonal components of the matrix self-energy determined with the Green's function \eqref{green} read (below $z=vp_z$):
\begin{eqnarray}
\Sigma_{1}(\varepsilon)&=&A \sum_{n\geq0}
\int_{-\infty}^{\infty}\!\!dz\ \frac{  \varepsilon-\Sigma_2+z}{(\varepsilon-\Sigma_1-z)(\varepsilon-\Sigma_2+z)-\Omega^2 n},
\nonumber
\\
\label{Sigma1SCBA}
\\
\Sigma_{2}(\varepsilon)&=&A \sum_{n\geq1}
\int_{-\infty}^{\infty}\!\! dz\ \frac{  \varepsilon-\Sigma_1-z}{(\varepsilon-\Sigma_1-z)(\varepsilon-\Sigma_2+z)-\Omega^2 n}.
\nonumber
\\
\label{Sigma2SCBA}
\end{eqnarray}
Here we introduced the energy scale $A$,
\begin{align}
A=\frac{\gamma\Omega^2}{8\pi v^3},
\end{align}
that combines the disorder coupling $\gamma$ and the strength of magnetic field
characterized by the distance $\Omega$ between the zeroth and first LLs.
In general, the self-energy depends on energy and on the LL index, $\hat{\Sigma}=\hat{\Sigma}(\varepsilon,p_z, n)$.
However, for the white-noise disorder, the dependences on $n$ and $p_z$ drop out.

For energies close to the Weyl point, $|\varepsilon|<\Omega$, and for weak disorder, $\beta\ll 1$,
the asymmetry with respect to the zeroth LL should be taken into account.
When the lowest LL is well separated from the others, $\text{Im}\Sigma_{1,2}<\Omega$,
the contribution of the sum over $n$ is dominated by the $n=0$ term. In this case, we get
\begin{align}\label{SCBA0}
\text{Im}\Sigma_1=-A\quad \text{ and }\quad \text{Im}\Sigma_2\sim-A\beta
\end{align}
Thus,  $\text{Im}\Sigma_2$ is negligible in the limit of weak disorder.

\begin{figure}
\includegraphics[width=\linewidth]{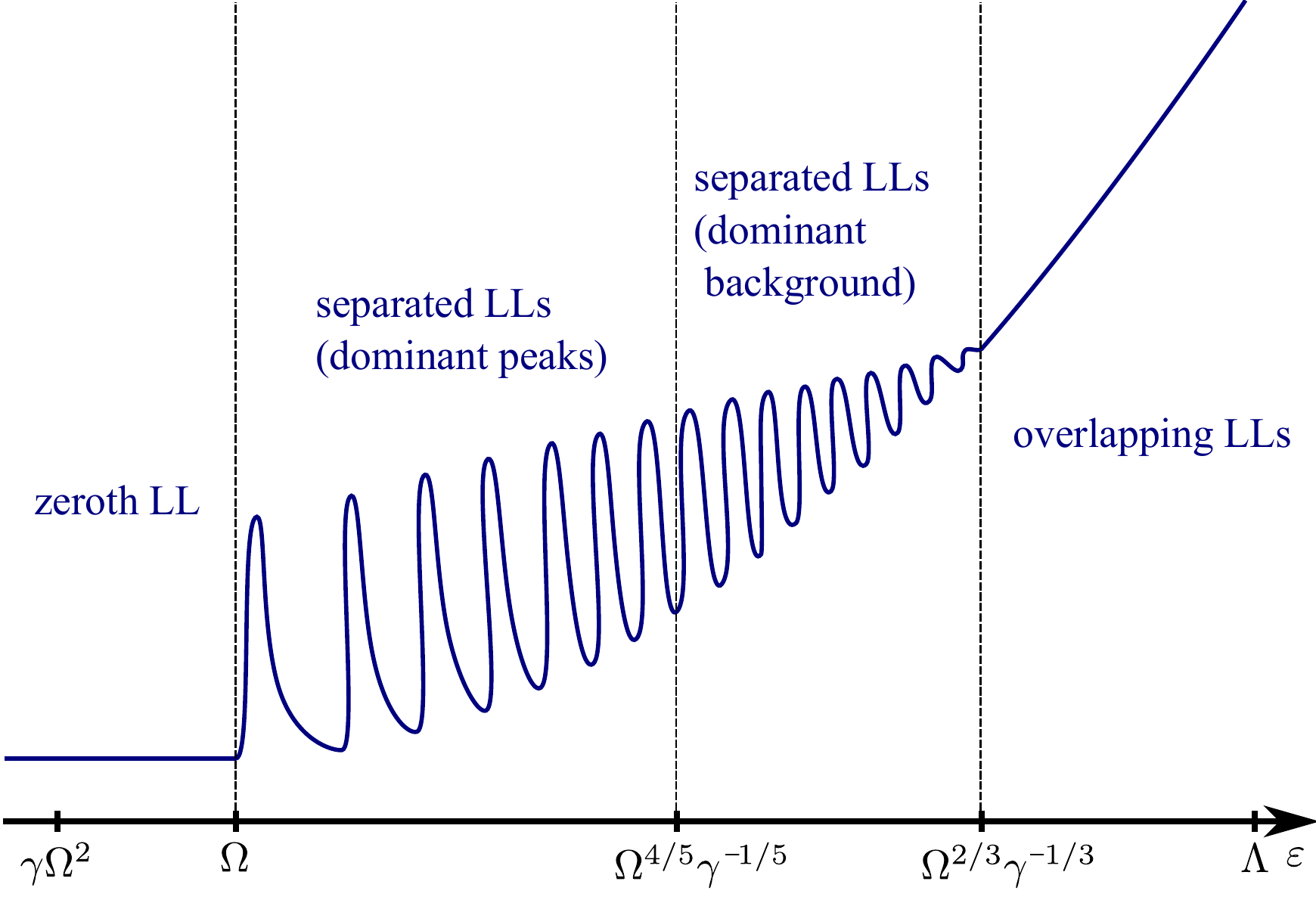}
\caption{Schematic plot of the density of states of a disordered Weyl semimetal in magnetic field. 
The shown energy scales indicate the borders of the different regimes: the zeroth LL ($\varepsilon<\Omega$), 
separated LLs ($\Omega<\varepsilon<\varepsilon_*$), separated LLs with dominating background ($\varepsilon_*<\varepsilon<\varepsilon_{**}$), 
and overlapping LLs ($\varepsilon_{**}<\varepsilon<\Lambda$). Here, $\varepsilon_*\sim\Omega(\Omega/A)^{1/5}\sim \Omega^{4/5}\gamma^{-1/5}$ 
and $\varepsilon_{**}\sim\Omega(\Omega/A)^{1/3}\sim \Omega^{2/3}\gamma^{-1/3}$ (for compactness, we set $v=1$ here and in further figures). 
For later purposes, within the range of zeroth LL we indicate the
scale $\gamma\Omega^2\ll \Omega$, at which the Hall conductivity $\sigma_{xy}$ of a single Weyl fermion compares to the conductivity $\sigma_{xx}$.
}
\label{energyscales}
\end{figure}

For energies away from the Weyl point, $\varepsilon>\Omega$, and weak disorder, $\beta<1$,
the asymmetry induced by the zeroth LL is negligible:
$\text{Im}\Sigma_1=\text{Im}\Sigma_2=-\Gamma$, where the LL broadening is determined by the
self-consistent equation
\begin{eqnarray}
\Gamma(\varepsilon)&\simeq &
A\varepsilon \sum_n\frac{\sqrt{\varepsilon^2-W_n^2+\sqrt{(W_n^2-\varepsilon^2)^2+4 \varepsilon^2 \Gamma^2(\varepsilon)}}}{\sqrt{2}\  \sqrt{(W_n^2-\varepsilon^2)^2+4 \varepsilon^2 \Gamma^2(\varepsilon)} }
\nonumber
\\
&=&\sum_{n}\Gamma_n
\label{GammanAPP}
\end{eqnarray}
with
$$W_n^2=\varepsilon_n^2+\Gamma^2(\varepsilon), \quad \varepsilon_n=\Omega^2n.$$
The solution of Eq.~\eqref{GammanAPP} gives a nonsymmetric peak of $\Gamma(\varepsilon)$ around the $n$th LL
located at $\varepsilon=W_n$ with
\begin{align}\label{SCBAN}
\Gamma(\varepsilon=W_n)\simeq
\begin{cases}
(A/2)^{2/3}\varepsilon^{1/3}, & \Omega\ll \varepsilon \ll \varepsilon_*,
\\
2 A (\varepsilon/\Omega)^2, & \varepsilon \gg \varepsilon_*,
\end{cases}
\end{align}
where $\varepsilon_*\sim\Omega(\Omega/A)^{1/5}$ marks the energy below which the LLs are fully separated.
A detailed analysis of the broadening of LLs reveals that the LLs are separated up to $\varepsilon_{**}\sim\Omega(\Omega/A)^{1/3}$,
but for energies in the range $\varepsilon_*<\varepsilon<\varepsilon_{**}$ the background density of states is larger than the density of states
for the particular LL as shown in Fig.~\ref{energyscales} (for further details, see Ref.~\cite{PhysRevB.92.205113}).

\section{Conductivity away from charge neutrality}
\label{sec:Sxx}

Using the introduced model, we calculate now the conductivity $\sigma_{xx}$ of a disordered Weyl semimetal
in the presence of magnetic field.
We restrict ourselves to the case of weak disorder, $\beta\ll 1$.
With the use of Kubo formula, the real part of the conductivity reads
\begin{align}\label{Kubo}
&\sigma_{xx}(\omega,T)=\int\frac{d\varepsilon}{2\pi}\frac{f_{T}(\varepsilon)}{\omega}\int\frac{d^{3}\textbf{p}}{(2\pi)^{3}}\nonumber\\
&\quad \times \text{Tr}\left\{\left[\hat{G}^{R}(\varepsilon,\textbf{p})-\hat{G}^{A}(\varepsilon,\textbf{p})\right]\hat{j}_{x}^\text{tr}
\hat{G}^{A}(\varepsilon-\omega,\textbf{p})\hat{j}_{x}\right.\nonumber\\
&\quad \left.+\hat{G}^{R}(\varepsilon+\omega,\textbf{p})\hat{j}_{x}^\text{tr}\left[\hat{G}^{R}(\varepsilon,\textbf{p})
-\hat{G}^{A}(\varepsilon,\textbf{p})\right]\hat{j}_{x}\right\},
\end{align}
where $\hat{j}_x=ev\sigma_x$ is the bare current operator and $\hat{j}_x^\text{tr}= V^\text{tr} \hat{j}_x$
is the current vertex dressed by disorder, see Ref.~\cite{PhysRevB.92.205113}.
We first calculate the conductivity without vertex corrections and include them at the final steps of the calculation.

After the evaluation of the trace and using the orthogonality of the wave functions of the different LLs,
Eq.~\eqref{Kubo} transforms into
\begin{align}
\sigma_{xx}^{(0)}\left(T\right)&=\frac{e^{2}v^{2}}{T}\int\frac{d\varepsilon}{2\pi}\frac{1}{\cosh^{2}\left(\frac{\varepsilon-\mu}{2T}\right)}
\  \sum_{n} \frac{eH}{2\pi c}
\notag
\\
&\times
\int\frac{dp_{z}}{2\pi}\ \text{Im} G^{R}_{11}(\varepsilon,n,p_{z})\ \text{Im}G^{R}_{22}(\varepsilon,n,p_{z}).
\label{transversal}
\end{align}
The Green functions here are written in the LL representation. We distinguish in the following calculations between the zeroth LL and higher LLs because the self-energies for the zeroth LL differ from those of the others. In the following, we will focus on low temperatures, $T\to 0$. For small chemical potential, $\mu< \Omega$, excitations to higher LLs are exponentially suppressed and the conductivity is dominated by the contribution of the zeroth LL. Note the conductivity in both region match via a narrow window at $\Omega\simeq\mu$ corresponding to the width of the first LL [cf. the last two lines in Eq.\eqref{consum}]. In the opposite regime, the conductivity is determined by the position of the chemical potential with respect to separated and overlapping LLs.

\subsection{Small chemical potential, $\mu<\Omega$: Zeroth Landau level}
\label{sub:zero}

We consider first the situation when the zeroth LL gives the dominant contribution to the conductivity.
This case is realized under the following two conditions: (i) the zeroth LL is separated from the first
one, which is fulfilled under the condition $A\ll \Omega$; (ii) the chemical potential satisfies $\mu< \Omega$,
while the temperature is close to zero, $T\to 0$. Under these conditions, the current vertex corrections are small,
$V^\text{tr}(\varepsilon\ll\Omega)\sim A/\Omega\ll 1$, for energies close to the Weyl node.
Therefore, we can disregard the difference between quantum and scattering time in the regime of the dominant zeroth LL contribution.
The Green function, using $\text{Im}\Sigma_1\simeq A$ and $\text{Im}\Sigma_1\simeq 0$ and disregarding the real parts of self-energies
($\text{Re}\Sigma\sim \beta\varepsilon\ll \varepsilon$, see Ref.~\cite{PhysRevB.92.205113}), reads
\begin{eqnarray}
G^{R}_{11}(\varepsilon,n,p_{z})&\simeq& \frac{\varepsilon+v p_z}{(\varepsilon+i A-v p_z)(\varepsilon+v p_z)-\Omega^2 n},
\label{G11e0}
\\
G^{R}_{22}(\varepsilon,n,p_{z})&\simeq& \frac{\varepsilon+iA-v p_z}{(\varepsilon+i A-v p_z)(\varepsilon+v p_z)-\Omega^2 (n+1)}.
\nonumber
\\
\label{G22e0}
\end{eqnarray}

Substituting Eqs.~\eqref{G11e0} and \eqref{G22e0} in Eq.~\eqref{transversal} and separating the $n=0$ term in the sum over all LLs,
we get
\begin{align}
\sigma_{xx}&=\frac{e^2\Omega^4A^2}{(2\pi)^2v}\int d\varepsilon
\frac{\partial f(\varepsilon)}{\partial\varepsilon}\int \frac{dz}{2\pi}\left\lbrace
\frac{1}{\left[(\varepsilon-z)^2+A^2\right]}\right.\nonumber\\
\times& \left.\frac{1}{\left[(\varepsilon^2-z^2-\Omega^2)^2+A^2(\varepsilon+z)^2\right]}\right.\nonumber\\
+&\left.\!\sum_{n=1}^{n_\text{max}}\frac{(\varepsilon+z)^2(n+1)}{\left[(\varepsilon^2-z^2-\Omega^2n)^2\right]
\left[(\varepsilon^2-z^2-\Omega^2(n+1))^2\right]}\right\rbrace,
\end{align}
where $n_\text{max}$ is the number of LLs within the energy band $\Lambda$.
After the integration over $\varepsilon$ for $T=0$, we find that the contribution of higher LLs is of the
order $e^2A^2/(\Omega v)$ and therefore negligible compared to the $n=0$ term that is of the order of $e^2A/v$.
For the dominant term coming from the zeroth LL we find
\begin{align}
\sigma_{xx}=&\frac{e^2\Omega^4A^2}{(2\pi)^2v}\int \frac{dz}{2\pi}\frac{1}{\left[(\mu-z)^2+A^2\right]\left[(z^2+\Omega^2)^2\right]}
\notag
\\
\simeq&\frac{e^2A}{(2\pi)^2v}.
\label{con1}
\end{align}
The result is proportional to the magnetic field and disorder strength and is equal
to the result of $\mu=0$, see Ref. \cite{PhysRevB.92.205113}.
A finite but small chemical potential, $\mu< \Omega$, does not essentially affect $\sigma_{xx}$:
the corrections to Eq.~(\ref{con1}) are small in the parameter $A\mu^2/\Omega^3$.

\subsection{Large chemical potential, $\mu>\Omega$}
\label{sub:allLL}

For large chemical potentials, $\mu>\Omega$, the situation is more subtle.
For a given magnetic field, the spectrum is subdivided in three domains: (i) the low-energy
part of the spectrum consists of separated LLs, (ii) in the intermediate region LLs are separated,
but the background density of states is larger than the height of an individual LL,
and, finally, (iii) at higher energies  the LLs overlap.
At low temperatures, the conductivity will strongly depend on the position of the chemical potential, with the unusual broadening of LLs
leading to an unconventional shape of the SdHO.

In view of the structure of the spectrum discussed above,  we need to distinguish for the calculation of the conductivity between the three different cases of the position of the chemical potential: (i) fully separated LLs, (ii) separated LLs, but large background, and (iii) fully overlapping LLs.
In all three cases the difference between the self-energies can be neglected and the self-energy can be written in terms of LL broadening: $\text{Im}\Sigma_1=\text{Im}\Sigma_2=-i\Gamma$.
The Green functions take then the form
\begin{eqnarray}
G^{R}_{11}(\varepsilon,n,p_{z})&\simeq& \frac{\varepsilon+v p_z+i\Gamma}{(\varepsilon+i \Gamma)^2-v^2 p_z^2-\Omega^2 n},
\label{G11n}
\\
G^{R}_{22}(\varepsilon,n,p_{z})&\simeq& \frac{\varepsilon+v p_z+i\Gamma}{(\varepsilon+i \Gamma)^2-v^2 p_z^2-\Omega^2 (n+1)},
\label{G22n}
\end{eqnarray}
Substituting these Green functions in the formula for the conductivity \eqref{transversal}, we perform the summation over $n$ and integration over $p_z$. (This calculation is analogous to that in the case $T\gg \Omega$ in Ref.~\cite{PhysRevB.92.205113}.)
The result is given by
\begin{align}\label{concalc}
\sigma_{xx}=\frac{e^2\Omega^2}{2\pi^2v}&\int d\varepsilon \frac{df(\varepsilon)}{d\varepsilon}\frac{2\Gamma\varepsilon^4}{\Omega^2\left[\Omega^4+(4\varepsilon\Gamma)^2\right]}\notag\\
&\times\left[\frac{4}{3}+\frac{\Omega^2}{\varepsilon}\left(\frac{\Gamma}{A\varepsilon}-\frac{2\varepsilon}{\Omega^2}\right)\right].
\end{align}

In all three cases of the structure of the spectrum near the chemical potential, the conductivity can be expressed by the semiclassical Drude formula, yielding
\begin{equation}
\sigma_{xx}^D=\frac{e^2v^2}{6\pi}
\int \frac{d\varepsilon}{4T \cosh^{2}\left(\frac{\varepsilon-\mu}{2T}\right)}\
\frac{\nu(\varepsilon)\tau_\text{tr}(\varepsilon)}{1+\omega_c^2(\varepsilon)[\tau_\text{tr}(\varepsilon)]^2}.
\label{Drudexx}
\end{equation}
Here $\tau_\text{tr}(\varepsilon)$ is the transport scattering time that takes into account the vertex corrections in $j_x^\text{tr}$
and is related to the quantum time $\tau_q=(2\Gamma)^{-1}$ via $\tau_\text{tr}=(3/2)\tau_q$.
In the case of overlapping LLs or of large background density of states compared to the particular LL,
$\varepsilon\gg \varepsilon^\ast$, the LL broadening is given by
\begin{equation}\label{tau_trans_back}
\Gamma(\varepsilon)=2A\frac{\varepsilon^2}{\Omega^2}=\frac{3}{4\tau_\text{tr}(\varepsilon)}.
\end{equation}
Using the SCBA relation between the density of states and the scattering time
\begin{equation}\label{DoSback}
\nu(\varepsilon)\tau_\text{tr}(\varepsilon)=\frac{3}{4\pi\gamma}
\end{equation}
and the semiclassical expression for the cyclotron frequency in the linear spectrum
\begin{equation}\label{cyclo}
\omega_c(\varepsilon)=\frac{v^2}{l_H^2\varepsilon}=\frac{\Omega^2}{2\varepsilon},
\end{equation}
we find the conductivity in this region:
\begin{eqnarray}
\sigma_{xx}\simeq
\frac{e^2}{\pi^2}\, \frac{A \Omega^2}{ v T }\!
\int \frac{d\varepsilon}{ \cosh^{2}\left(\frac{\varepsilon-\mu}{2T}\right)}
\frac{\varepsilon^6}{(8A\varepsilon^3)^2+9\Omega^8/4}.
\label{sigmaxx-class}
\end{eqnarray}
In the following, we use  Eq.~(\ref{sigmaxx-class}) to  evaluate the conductivity in all three regimes.

First, we consider  the regime of fully separated LLs, when the relevant energies satisfy $\Omega\ll \varepsilon \ll \Omega(\Omega/A)^{1/5}$, assuming that the chemical potential is located within one of the LLs and the temperature is low (smaller than the LL width).
The conductivity for a general LL broadening is given by
\begin{align}\label{consep}
\sigma_{xx}&\simeq \frac{e^2\Omega^2}{\pi^2vA}\int_{-\infty}^\infty d\varepsilon\delta(\varepsilon-\mu)\frac{4\Gamma^2\varepsilon^2}{9\Omega^4}\nonumber\\
&\simeq\frac{4e^2\mu^2\Gamma^2}{9\pi^2vA\Omega^2}.
\end{align}
The broadening of the LLs at the LL center  is given by $\Gamma=(A/2)^{2/3}\varepsilon^{1/3}$, which yields the conductivity in the center of LLs (in the following denoted by $\sigma^\text{peak}_{xx}$)
\begin{align}\label{conpeak}
\sigma^\text{peak}_{xx}\sim\frac{\gamma^{1/3}\mu^{8/3}}{\Omega^{4/3}}.
\end{align}
The conductivity of the background density of states with a broadening of $\Gamma\sim\gamma\varepsilon^2$ is denoted by $\sigma_{xx}^\text{bg}$ and reads
\begin{align}
\sigma_{xx}^\text{bg}\simeq \frac{2e^2\gamma \mu^6}{9\pi^3v^4\Omega^4}.
\end{align}

Next, we turn to the intermediate range of the location of the chemical potential, $\Omega(\Omega/A)^{1/5}\ll \epsilon \ll \Omega(\Omega/A)^{1/3}$. In this case, the $\Omega^8$-term in the denominator of Eq.~(\ref{sigmaxx-class}) dominates, yielding
\begin{align}
\label{conback}
&\sigma_{xx}\simeq \frac{4e^2A\Omega^2}{9\pi^2vT}\int\frac{d\varepsilon}{\cosh\left(\frac{\varepsilon-\mu}{2T}\right)}\frac{\varepsilon^6}{\Omega^8}\nonumber\\
&\ =\frac{2e^2\gamma}{9\pi^3v^4\Omega^4}\left(\mu^6+5\pi^2\mu^4T^2+7\pi^4\mu^2T^4+\frac{31}{21}\pi^6T^6\right) \nonumber \\
&\  \simeq \frac{2e^2\gamma \mu^6}{9\pi^3v^4\Omega^4}.
\end{align}
In the last line of Eq.~(\ref{conback}) we have taken low-temperature limit (here the condition $T\ll \mu$ is sufficient).

Finally, for higher chemical potential, $\varepsilon>\Omega(\Omega/A)^{1/3}$, which is the regime of overlapping LLs, we neglect $\Omega^8$ in the denominator of Eq.~(\ref{sigmaxx-class}), which leads to
\begin{align}\label{conover}
\sigma_{xx}&=\frac{e^2A\Omega^2}{\pi^2vT}\int\frac{d\varepsilon}{\cosh^2\left(\frac{\varepsilon-\mu}{2T}\right)}
\frac{\varepsilon^6}{(8A\varepsilon^3)^2}\nonumber\\
&=\frac{e^2v^2}{2\pi\gamma} .
\end{align}
The result coincides with the conductivity $\sigma_{xx,0}$ in the absence of magnetic field and does not depend on the chemical potential.

Magnetooscillations of the conductivity stem from the oscillations of the density of states $\nu(\varepsilon)$ and of the transport scattering time $\tau_\text{tr}(\varepsilon)$, see Ref.~\cite{RevModPhys.84.1709}.
For a Weyl semimetal,  the density of states with magnetooscillations is given by
\begin{align}\label{muosc}
\nu(\varepsilon)=&\,\nu_0\, \left\{1+\sum_{k=1}^\infty\sqrt{\frac{\omega_c(\varepsilon)}{2k\varepsilon}}\,
\delta^k\right.
\nonumber\\
&\quad \times\left.\left[\cos\frac{\pi k\varepsilon}{\omega_c(\varepsilon)}
+\sin\frac{\pi k\varepsilon}{\omega_c(\varepsilon)}\right]\right\},
\end{align}
where 
$$\delta=\exp\left[-\frac{\pi}{\omega_c(\varepsilon)\tau_{q}(\varepsilon)}\right]$$ 
is the Dingle factor
determined by the quantum scattering time $\tau_q$.
Note that in the case of a conventional 3D material with parabolic dispersion (see Ref.~\cite{Vasko}), the frequency of the oscillations
is a factor of $2$ larger then in the case of Weyl semimetals. A similar behavior is encountered in the 2D case of graphene \cite{PhysRevB.87.195432}
in comparison to conventional 2D materials. The non-equidistant behavior of the LLs for relativistic dispersion relations
is expressed via the energy dependent cyclotron frequency $\omega_c$.
For $\omega_c(\varepsilon)\tau_q(\varepsilon)\ll 1$, which corresponds exactly to the condition of overlapping LLs,
the first harmonics, $k=1$, is the least damped term and hence dominates the oscillations.

Using Eqs.~(\ref{muosc}) and (\ref{DoSback}), we find the oscillatory contribution to the conductivity (the SdHO) for the case of overlapping LLs:
\begin{align}
\sigma_{xx}\simeq &\sigma_{xx,0}\left\{1+\frac{3\gamma\mu^2}{2\pi v^3\Omega}\exp\left(-\frac{\gamma\mu^3}{v^3\Omega^2}\right)
\right.\nonumber\\
&\left.\qquad \times
\left[\cos\left(\frac{2\pi\mu^2}{\Omega^2}\right)+\sin\left(\frac{2\pi\mu^2}{\Omega^2}\right)\right]\right\},
\end{align}
where $\sigma_{xx,0}$ is the smooth part of the conductivity calculated above [Eq.~\eqref{conover}].
As usual, the SdHO are exponentially damped in the regime of overlapping LLs, in contrast to the case of separated LLs.

We conclude this section with a summary of the results for the conductivity,
\begin{align}\label{consum}
\sigma_{xx}\!=\!\frac{e^2}{\pi^2v}\left\lbrace\begin{array}{ll}
\!\dfrac{v^3\pi}{2\gamma}, &\ \Omega\ll \mu^{3/2}\gamma^{1/2}\\[0.4cm]
\!\dfrac{2\gamma\mu^6}{9\pi v^3\Omega^4}, &\ \mu^{3/2}\gamma^{1/2}\ll\!\Omega\!\ll\!\mu^{5/4}\gamma^{1/4}\!,\\[0.4cm]
\!\dfrac{32\pi v^3\mu^2\Gamma^2(\mu)}{9\gamma\Omega^4}, &\ \mu^{5/4}\gamma^{1/4}\ll \Omega< \mu,\\[0.4cm]
\!\dfrac{\gamma\Omega^2}{32\pi v^3}, &\ \mu < \Omega \ll \gamma^{-1}
\end{array}\right.
\end{align}
in the different regimes with respect to magnetic field, chemical potential, and disorder strength.

\section{Hall conductivity}
\label{sec:Hall}

In this section, we calculate the Hall conductivity. According to the Kubo-Streda formula~\cite{0022-3719-15-22-005}, the Hall conductivity is given by
\begin{align}\label{Stredadef}
\sigma_{xy}\!=\!\frac{i e^2}{2\pi}\int\! d\varepsilon f(\varepsilon)\text{Tr}\left[v_x\frac{dG^\text{R}}{d\varepsilon}v_y\text{Im}G-v_x\text{Im}Gv_y\frac{dG^A}{d\varepsilon}\right]\!,
\end{align}
It is convenient to split up the Hall conductivity into a normal, $\sigma_{xy}^\text{I}$, and an anomalous, $\sigma_{xy}^\text{II}$, contributions.
The normal contribution is determined by states near the Fermi level and
can be simplified by using the orthogonality of the wave functions of different LLs. We find
\begin{align}\label{normHall}
\sigma_{xy}^\text{I}=&\frac{e^2\Omega^2}{(2\pi)^2}\int d\varepsilon \frac{df(\varepsilon)}{d\varepsilon}\int\frac{dp_z}{2\pi}\sum_{n}\left[G^\text{R}_{22}\text{Im}G_{11}\right.\nonumber\\
&\left.-G^\text{R}_{11}\text{Im}G_{22}-\text{Im}G_{22}G^\text{A}_{11}+
\text{Im}G_{11}G^\text{A}_{22}\right].
\end{align}
The anomalous contribution reflects the thermodynamic properties of the system in the presence of magnetic field
and can be expressed as
\begin{align}\label{Stredanu}
\sigma_{xy}^\text{II}=e\frac{\partial N(H,\mu)}{\partial H}.
\end{align}
Here $N$ is the electron density defined as follows:
\begin{align}\label{partdens}
N(H,\mu)=\frac{1}{V}\sum_{\vec{p}}f(\varepsilon_{\vec{p}})=\int_{-\infty}^\infty d\varepsilon\frac{\nu(\varepsilon)}{\exp\left(\frac{\varepsilon-\mu}{T}\right)+1}.
\end{align}
Below, we will first calculate the Hall conductivity in the clean case, and then will incorporate disorder which is encoded in the density of states $\nu(\varepsilon)$.

\subsection{Clean case}

\begin{figure}
\begin{center}
\includegraphics[scale=0.6]{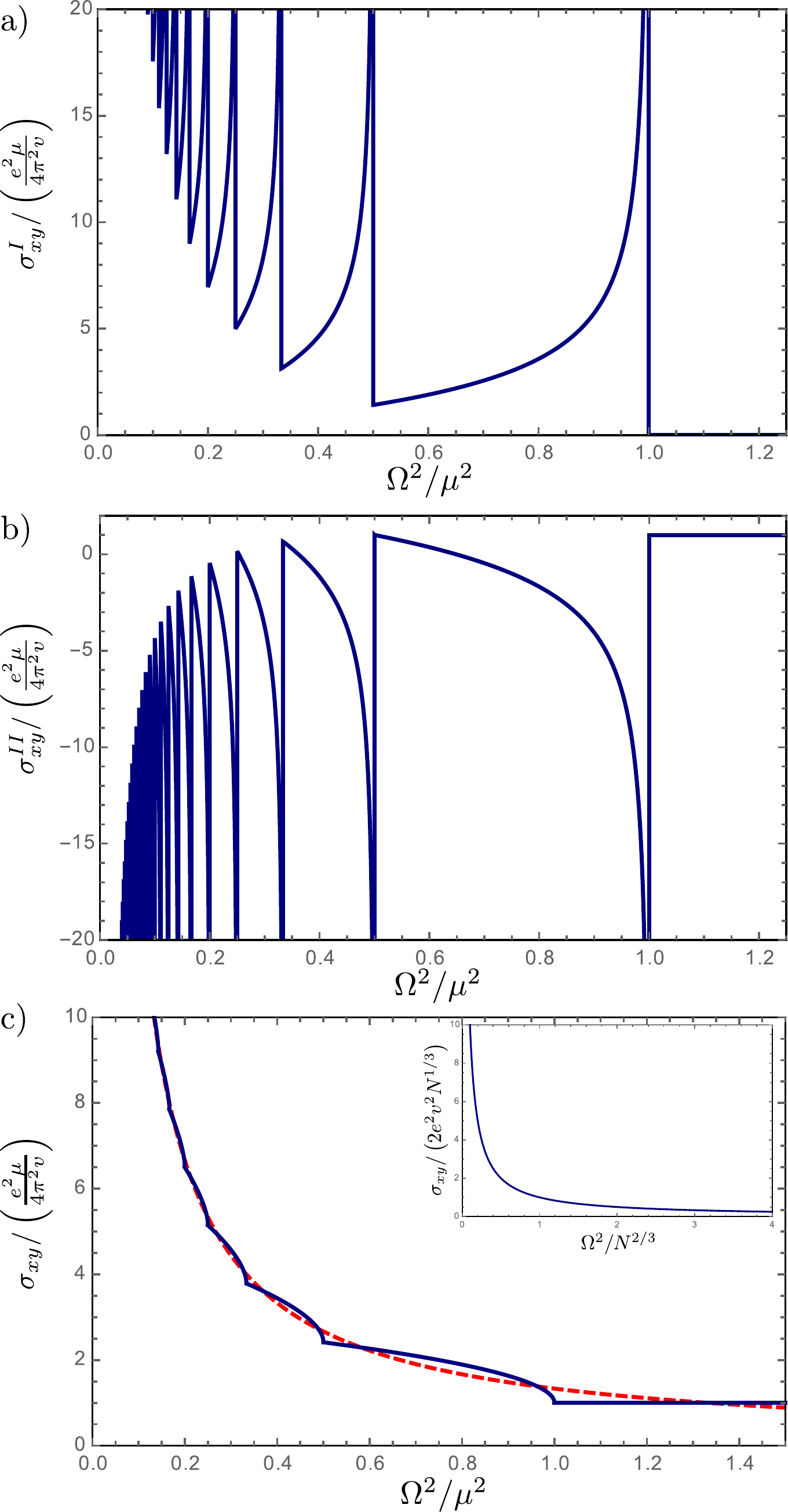}
\end{center}
\caption{Normal part of the Hall conductivity (a), anomalous part (b), and the total Hall conductivity (c) of a clean Weyl semimetal. 
All three curves are plotted as functions of magnetic field ($\Omega^2\propto H$) for a fixed chemical potential. 
In panel (c) the red dashed curve corresponds to the smoothened Hall conductivity (with SdHO subtracted).
The inset in panel (c) visualizes the Hall conductivity for a fixed particle density.}
\label{fig:Hall_clean}
\end{figure}

We now briefly discuss the Hall conductivity in the clean case.
The Green functions in Landau representation for the clean case read
\begin{align}
G_{11}(\varepsilon, p_z,n)&=\sum_{\lambda}\left(1+\frac{\lambda vp_z}{\varepsilon_n}\right)
\frac{1}{\varepsilon-\lambda\varepsilon_n+i0},
\label{cleanG1}\\
G_{22}(\varepsilon, p_z,n)&=\sum_{\lambda}\left(1-\frac{\lambda vp_z}{\varepsilon_{n+1}}\right)
\frac{1}{\varepsilon-\lambda\varepsilon_{n+1}+i0},
\label{cleanG2}
\end{align}
We start with the calculation of the normal part of Hall conductivity and substitute the Green
function from Eqs.~\eqref{cleanG1} and \eqref{cleanG2} in Eq.~\eqref{normHall}.
After the evaluation of the integral over energy $\varepsilon$ and of sum over energy bands $\lambda$,
the normal contribution to the Hall conductivity reads
\begin{align}\label{normHallcalc}
\sigma_{xy}^\text{I}&=-\frac{e^2\Omega^2}{4\pi T}\int\frac{dp_z}{2\pi}\sum_{n=1}^{\infty}\text{Re}\left(\frac{n}{\varepsilon_n}\right.
\nonumber\\
&
\times \left.\left[\frac{1}{\cosh^2\left(\frac{\varepsilon_n+\mu}{2T}\right)}-\frac{1}{\cosh^2\left(\frac{\varepsilon_n-\mu}{2T}\right)}\right]\right).
\end{align}
The evaluation of the integrals for $T=0$ leads to
\begin{align}
\sigma_{xy}^\text{I}=\frac{e^2\Omega^2}{2\pi^2 v}\sum_{n=1}^{(\mu/\Omega)^2}\frac{n}{\sqrt{\mu^2-\Omega^2n}}.
\label{sigmaIxy-clean}
\end{align}
The normal contribution of the Hall conductivity shows singularities when the chemical potential is at the center of the one particular LL, $\mu=\Omega\sqrt{n}$, see Fig. ~\ref{fig:Hall_clean} (a).

The anomalous contribution to the Hall conductivity is
obtained from Eq.~\eqref{Stredanu} and the density of states $\nu(\varepsilon)$ of a Weyl semimetal in clean case,
\begin{align}
\nu(\varepsilon)=\frac{1}{4\pi^2l^2v}\left[1+2\sum_{n=1}^{\varepsilon^2l^2/2}\frac{|\varepsilon|}{\sqrt{\varepsilon^2-\Omega^2n}}\right].
\end{align}
We evaluate the integral in Eq.~(\ref{partdens}) for $T=0$ and take the derivative of $N$ with respect to magnetic field $H$:
\begin{align}\label{anhall}
\sigma_{xy}^\text{II}&=-\frac{e^2\Omega^2}{2\pi^2 v}\sum_{n=1}^{\mu^2/\Omega^2}\frac{n}{\sqrt{\mu^2-\Omega^2n}}\nonumber\\
&+\frac{e^2}{4\pi^2 v}\left[\mu+2\sum_{n=1}^{\mu^2/\Omega^2}\sqrt{\mu^2-\Omega^2n}\right]
\end{align}
The first term of Eq.~\eqref{anhall} also shows singularities when the chemical potential is at the center of the one particular LL opposite of those of the normal contribution from Eq.~(\ref{sigmaIxy-clean}), see Fig.~\ref{fig:Hall_clean} (b). Therefore, these singularities are exactly canceled in the total Hall conductivity.
As demonstrated in Appendix~\ref{app:anhall}, this cancellation occurs in the clean case in the general case of arbitrary $T$.

The evaluation of the sum over LLs with Euler-Maclaurin formula leads to the leading order to
\begin{equation}\label{sigma_xy_mu}
\sigma_{xy}\simeq \frac{e^2}{4\pi^2 v}\left\{
\begin{array}{ll}
\mu, & \mu<\Omega, \\[0.2cm]
\dfrac{4\mu^3}{3\Omega^2}, & \mu>\Omega.\\[0.2cm]
\end{array}\right.
\end{equation}
For $\mu>\Omega$, Eq.~\eqref{sigma_xy_mu} describes the smoothened part of the Hall conductivity.
On top of this background contribution there is an oscillatory part induced by the Landau quantization.
The Hall conductivity (normal and anomalous part and the total Hall conductivity) without disorder is visualized in Fig.~\ref{fig:Hall_clean}, 
where the oscillations induced by Landau quantization
can be seen clearly in case of fixed chemical potential. Already based on this plot, one can expect that in the presence of disorder
the total Hall conductivity is only weakly changed, since the disorder-induced broadening would only smoothen the oscillatory part of the curve.

Further, we can express the Hall conductivity for a fixed particle density $N$ instead of a fixed chemical potential, as relevant to experiments.
The magneto-oscillations in the chemical potential are then exactly canceled by the oscillations in the particle density:
\begin{align}
\sigma_{xy}=2e^2v^2\frac{N}{\Omega^2},
\label{sigma-xy}
\end{align}
see inset in Fig.~\ref{fig:Hall_clean} (c).
Here the zero level of the density $N$ is chosen in such a way that $N=0$ for the chemical potential located in the Dirac point, $\mu=0$.

\subsection{Normal Hall conductivity in the presence of disorder}
\label{sub:normalHall}

Now, we turn to the Hall conductivity in the presence of disorder and first proceed with the evaluation of the normal contribution.
As explained in Sec.\ref{sec:model}, we distinguish again between the cases when the chemical potential is within the zeroth LL or higher LLs.
We focus on low temperatures, $T\to 0$,
throughout the whole section.

We will start with the calculation of the Hall conductivity under the following conditions:
(i) the zeroth LL is separated from higher LLs, $A\ll \Omega$;
(ii) excitations to higher LLs are suppressed, $\mu< \Omega$.
Using the Green functions for energies close to the zeroth LL, Eqs.~\eqref{G11e0} and \eqref{G22e0},
the formula for normal contribution to the Hall conductivity, Eq.~\eqref{normHall} transforms to
\begin{widetext}
\begin{align}
\sigma_{xy}^{\text{I}}=&-\frac{e^2\Omega^2A}{(2\pi)^2 v}\int \frac{dz}{2\pi}\sum_n\frac{A^2(\mu+z)^3+(\mu+z)^2(\mu-z)[\mu^2-z^2-\Omega^2(n+1)]-(\mu+z)(\mu^2-z^2-\Omega^2n)\Omega^2(n+1)}
{\left[(\mu^2-z^2-\Omega^2n)^2+A^2(\mu+z)^2\right]\left[(\mu^2-z^2-\Omega^2(n+1))^2+A^2(\mu+z)^2\right]},
\label{}
\end{align}
\end{widetext}
where $z=vp_z$.

In the following, we will split the summation over the LL index into the term with $n=0$ and the terms with $n>0$.
In contrast to the conductivity $\sigma_{xx}$, the contribution of the terms with $n>0$ in $\sigma_{xy}^{\text{I}}$  is of the same order as the $n=0$ term.
The evaluation of  the terms under the conditions $A\ll \Omega$ and $\mu< \Omega$ gives to the leading order
\begin{align}
\sigma_{xy}^{\text{I}}\simeq \frac{5e^2A\mu}{4(2\pi)^2v\Omega}\sim\frac{e^2}{v^4}\gamma\mu\Omega.
\label{sigma-xy-I-disordered}
\end{align}
Clearly, this result (linear in disorder) matches the result for a clean system,
where the normal contribution for the case of the chemical potential located in the zeroth Landau level is absent.
We will see below that the term (\ref{sigma-xy-I-disordered}) is negligible in comparison with the anomalous contribution to the Hall conductivity.

Now, we turn to higher chemical potential $\mu> \Omega$ and analyze the contribution of higher LLs to $\sigma_{xy}^{\text{I}}$.
For  $\varepsilon\gg \Omega$, the difference between the self-energies for the two bands can be neglected and we
can use the Green functions \eqref{G11n} and \eqref{G22n} in Eq.~\eqref{normHall}.
The detailed calculation is presented in Appendix~\ref{app:condhighmu}.
The normal contribution to the Hall conductivity reads
\begin{align}\label{Hallcalc}
\sigma_{xy}^{\text{I}}\simeq\frac{e^2\Omega^2}{2\pi^2v}&\int d\varepsilon \frac{df(\varepsilon)}{d\varepsilon}\frac{\varepsilon^3}{\Omega^4+(4\varepsilon\Gamma)^2}\nonumber\\
&\times\left[\frac{4}{3}+\frac{\Omega^2}{\varepsilon}\left(\frac{\Gamma}{A\varepsilon}-\frac{2\varepsilon}{\Omega^2}\right)\right].
\end{align}
The limit of vanishing disorder, $\Gamma\to0$, is reproduced in Eq.~\eqref{app:Hall_calc}.
Similarly to $\sigma_{xx}$, the normal contribution to the Hall conductivity can be cast in the form of a semiclassical Drude formula:
\begin{equation}
\sigma_{xy}^{\text{I,D}}\simeq\frac{e^2v^2}{6\pi}
\int \frac{d\varepsilon}{4T \cosh^{2}\left(\frac{\varepsilon-\mu}{2T}\right)}\
\frac{\nu(\varepsilon)\tau_\text{tr}(\varepsilon)\omega_c(\varepsilon)[\tau_\text{tr}(\varepsilon)]}{1+\omega_c^2(\varepsilon)[\tau_\text{tr}(\varepsilon)]^2}.
\label{Drudexy}
\end{equation}

In the regime where the contribution of the separated LLs to the density of states exceeds the contribution of the background,
$\Omega\ll \mu \ll \Omega(\Omega/A)^{1/5}$,
the second term in Eq.~\eqref{Hallcalc} dominates. In the limit $T\to 0$, we get
\begin{align}
\sigma_{xy}^{\text{I}}&=\frac{e^2\Omega^2}{2\pi^2v}\int d\varepsilon \delta(\varepsilon-\mu)\,\frac{2\Omega^2\varepsilon^2}{\Omega^4+(4\varepsilon\Gamma)^2}\frac{\Gamma}{A\varepsilon}\nonumber\\
&\simeq\frac{e^2\mu\Gamma(\mu)}{2\pi^2vA}.
\label{sigma-xy-I-separated}
\end{align}

Next, we evaluate the Hall conductivity for a larger chemical potential, when the LLs are separated but the contribution of the background dominates, or else, the LLs fully overlap. In these cases, the expressions for the density of states, transport scattering time, and cyclotron frequency are given by Eqs.~\eqref{tau_trans_back}, \eqref{DoSback}, and \eqref{cyclo}, respectively.
In the range $\Omega(\Omega/A)^{1/5}\ll \mu \ll \Omega(\Omega/A)^{1/3}$, which corresponds to the case of separated LLs with the dominant background density of states,
we find
\begin{align}
\sigma_{xy}^{\text{I}}&=\frac{e^2}{\pi^2v}\int d\varepsilon \frac{1}{4T\cosh^2\left(\frac{\varepsilon-\mu}{2T}\right)}\frac{\varepsilon^3}{\Omega^2}
\nonumber\\
&=\frac{e^2\mu^3}{\pi^2v\Omega^2}\left(1+\frac{\pi^2T^2}{\mu^2}\right).
\end{align}
For fully overlapping LLs, $\mu\gg \Omega(\Omega/A)^{1/3}$, the normal contribution to the Hall conductivity reads
\begin{align}\label{Hallover}
\sigma_{xy}^{\text{I}}&\simeq \frac{3e^2\Omega^2}{2\pi^2v}\int d\varepsilon \frac{1}{4T\cosh^2\left(\frac{\varepsilon-\mu}{2T}\right)}\frac{\pi^2v^5}{\gamma^2\varepsilon^3}\nonumber\\
&=\frac{3e^2v^5\Omega^2}{2\mu^3\gamma^2}.
\end{align}

\subsection{Anomalous Hall conductivity in the presence of disorder}
\label{sub:anomalHall}

In this Section, we calculate the anomalous contribution to the Hall conductivity in the presence of disorder.
Furthermore, we subtract the contribution of states below the charge neutrality point since they do not contribute
to the Hall conductivity.
This is shown explicitly in Appendix~\ref{app:anhall} for the clean case and holds for finite disorder
in the weak disorder regime, $\gamma\Lambda<1$, considered here.
The density of states of a disordered Weyl semimetal is given by
\begin{align}
\nu(\varepsilon)=-\frac{1}{\pi\gamma}\left(\text{Im}\Sigma_1+\text{Im}\Sigma_2\right).
\end{align}
In the calculation of the self-energy, we distinguish between the zero LL and the others.
For the energy at the zeroth LL the self-energy is given by
\begin{align}
\text{Im}\Sigma_1=-A\qquad\text{and}\qquad\text{Im}\Sigma_2=0,
\end{align}
which will be used in the regime $\mu< \Omega$.
The anomalous Hall conductivity in this regime does not depend on weak disorder, $\gamma\Lambda\ll 1$:
\begin{align}\label{Hall0}
\sigma_{xy}^\text{II}\simeq e\frac{\partial}{\partial H}\int^{\infty}_{0} d\varepsilon
\frac{1}{\exp\left(\frac{\varepsilon-\mu }{2T}\right)+1}\frac{A}{\pi\gamma}=\frac{e^2\mu }{4\pi^2 v},
\end{align}
This result matches the ac anomalous Hall conductivity $\sigma_{xy}(\omega)$ obtained in Ref.~\cite{2017arXiv170404258S}
in the limit $\omega \to 0$.

For $\mu>\Omega$ the situation is more subtle. The self-energy depends on the strength of broadening and, for separated LLs, $\mu<\Omega(\Omega/A)^{1/3}$, on the actual position of the chemical potential with respect to the center of a given LL. The shape of the density of states consists of the peak at the center of the LL, the tail of the LL, and the background, see Ref.~ \cite{PhysRevB.92.205113}. For separated LLs with large background and for overlapping LLs, the density of states is dominated by the background contribution. The anomalous Hall conductivity for $\mu> \Omega$ reads
\begin{align}
\sigma_{xy}^\text{II}=e\frac{\partial}{\partial H}\int_{0}^\mu d \varepsilon \frac{2\Gamma(\varepsilon)}{\pi\gamma},
\end{align}
where $\Gamma(\varepsilon)$ is given by Eq.~\eqref{GammanAPP}.

\begin{figure}
\begin{center}
\includegraphics[scale=0.6]{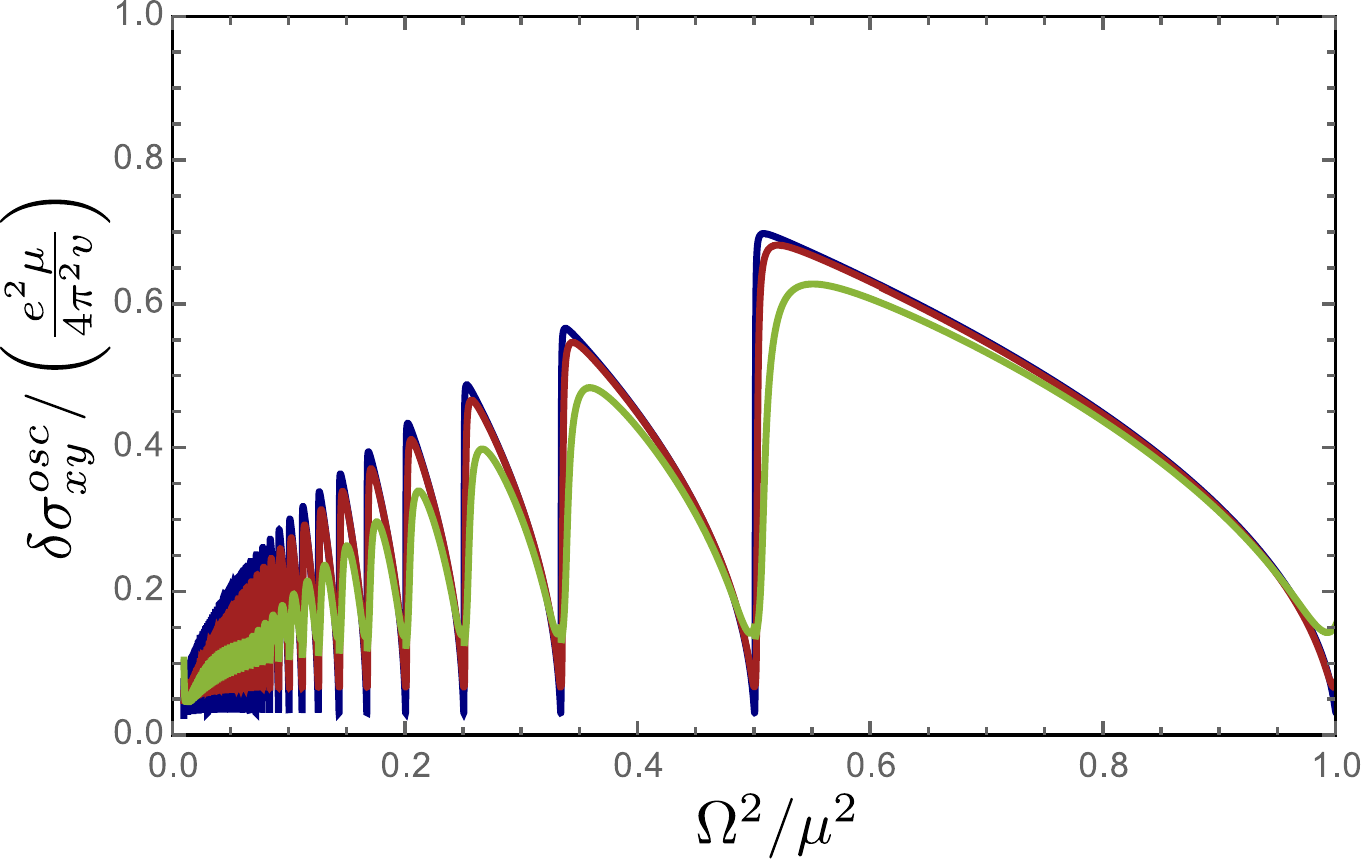}
\end{center}
\caption{The oscillatory part of the anomalous Hall conductivity in presence of disorder
as obtained by the numerical solution of the equation for the self-energy, Eq.\eqref{anoHalltail}.
Blue, red, and green curves correspond to $A\mu/\Omega^2=10^{-5},10^{-4},10^{-3}$, respectively.
The ultraviolet cutoff for all curves was set to $\Lambda/\Omega=100$.}
\label{Halldis}
\end{figure}

Under the same approximations as in the calculation of $\sigma_{xy}^\text{I}$, we obtain the anomalous Hall conductivity in the disordered case, reading
\begin{align}\label{anoHalltail}
\sigma_{xy}^\text{II}
&\simeq\sum_n\frac{e^2}{2\pi^2v}\frac{A\mu}{\Gamma_n(\mu)}-\frac{e^2\mu\Gamma(\mu)}{2\pi^2vA}\notag\\
&=\sum_n\frac{e^2}{2\pi^2v}\frac{A\mu}{\Gamma_n(\mu)}-\sigma_{xy}^\text{I},
\end{align}
where $\Gamma_n$ is defined in Eq.~\eqref{GammanAPP}. For $\Gamma\to 0$ in Eq.~\eqref{GammanAPP}, the result (\ref{anhall}) obtained in the limit of vanishing disorder is reproduced. Moreover, for non-overlapping LLs, the broadening of LLs in Eq.~(\ref{anoHalltail}) is only important 
in the term the sum over LLs that corresponds
to the LL where the chemical potential is located; for all other $n$ one can replace $A\mu/\Gamma_n(\mu)$ with $\sqrt{\mu^2-\Omega^2n}$, as in Eq.~(\ref{anhall}).
The smoothened part of the Hall conductivity for separated LLs $\mu^{3/2}\gamma^{1/2}\ll \Omega < \mu$,
\begin{align}\label{Hall}
\sigma_{xy}\simeq\frac{e^2}{4\pi^2 v}
\dfrac{4\mu^3}{3\Omega^2}
\end{align}
is thus the same as in the limit without disorder.  
The effects of the oscillations are minor compared to smoothened part of the Hall conductivity. 
Therefore, we will use Eq.~\eqref{Hall} in the following sections to calculate the magnetoresistance.
The oscillatory part of the Hall conductivity for fully separated LLs shown in Fig.~\ref{Halldis} visualizes the effect of disorder 
in the Hall conductivity. 

For overlapping LLs, the main term in the broadening is given by $\Gamma=2A\varepsilon^2/\Omega^2$ which is independent of magnetic field
and therefore the anomalous Hall conductivity is zero to the leading order.
The corrections due to magnetic field in the case of overlapping LLs are proportional to the Dingle factor,
as described above. The particle density for zero temperature reads
\begin{align}\label{Nover}
N(H,\mu)=&\int_0^\mu d\varepsilon\nu(\varepsilon)\left\{1+\sqrt{\frac{\omega_c(\varepsilon)}{2\varepsilon}}\right.
\delta\, \nonumber\\
&\times\left.\left[\cos\frac{\pi\varepsilon}{\omega_c(\varepsilon)}+\sin\frac{\pi\varepsilon}{\omega_c(\varepsilon)}\right]\right\}.
\end{align}
Since the Dingle factor is exponentially small for overlapping LLs, the anomalous part of the Hall
conductivity decays exponentially. The same applies for the TMR.
The contributions of overlapping LLs to the TMR will therefore be dominated by effects of finite
temperature and will not be discussed here.

\section{Magnetoresistance for pointlike impurities}
\label{sec:magneto}

We now turn to the evaluation of the TMR,
\begin{align}
\Delta_\rho(H)=\frac{\rho_{xx}(H)-\rho_{xx}(0)}{\rho_{xx}(0)},
\end{align}
which quantifies the difference between the resistivity $\rho_{xx}(H)$ in a finite magnetic field
and the resistivity at $H=0$.
Using
$$
\rho_{xx}=\frac{\sigma_{xx}}{\sigma_{xx}^2+\sigma_{xy}^2},
$$
we express the TMR through the conductivities at zero and finite magnetic fields, $\sigma_{xx}(0)$ and $\sigma_{xx}(H)$,
as well as the Hall conductivity $\sigma_{xy}(H)$,
\begin{align}\label{magres}
\Delta_\rho(H)=\frac{\sigma_{xx}(H)\sigma_{xx}(0)}{\sigma_{xx}^2(H)+\sigma_{xy}^2(H)}-1,
\end{align}
and employ the results from the previous sections.

The results for the TMR are either dominated by a large conductivity, $\sigma_{xx}\gg \sigma_{xy}$, leading to
\begin{align}\label{rho-by-xx}
\Delta_\rho(H)\simeq\frac{\sigma_{xx}(0)}{\sigma_{xx}(H)}-1
\end{align}
or dominated by a large Hall conductivity, $\sigma_{xy}\gg \sigma_{xx}$, resulting in
\begin{align}\label{rho-by-xy}
\Delta_\rho(H)\simeq\frac{\sigma_{xx}(H)\sigma_{xx}(0)}{\sigma_{xy}^2(H)}-1.
\end{align}
In what follows, we will distinguish between fixed chemical potential and fixed particle density. Let us start with fixed chemical potential $\mu$.

\begin{figure}
\includegraphics[width=\linewidth]{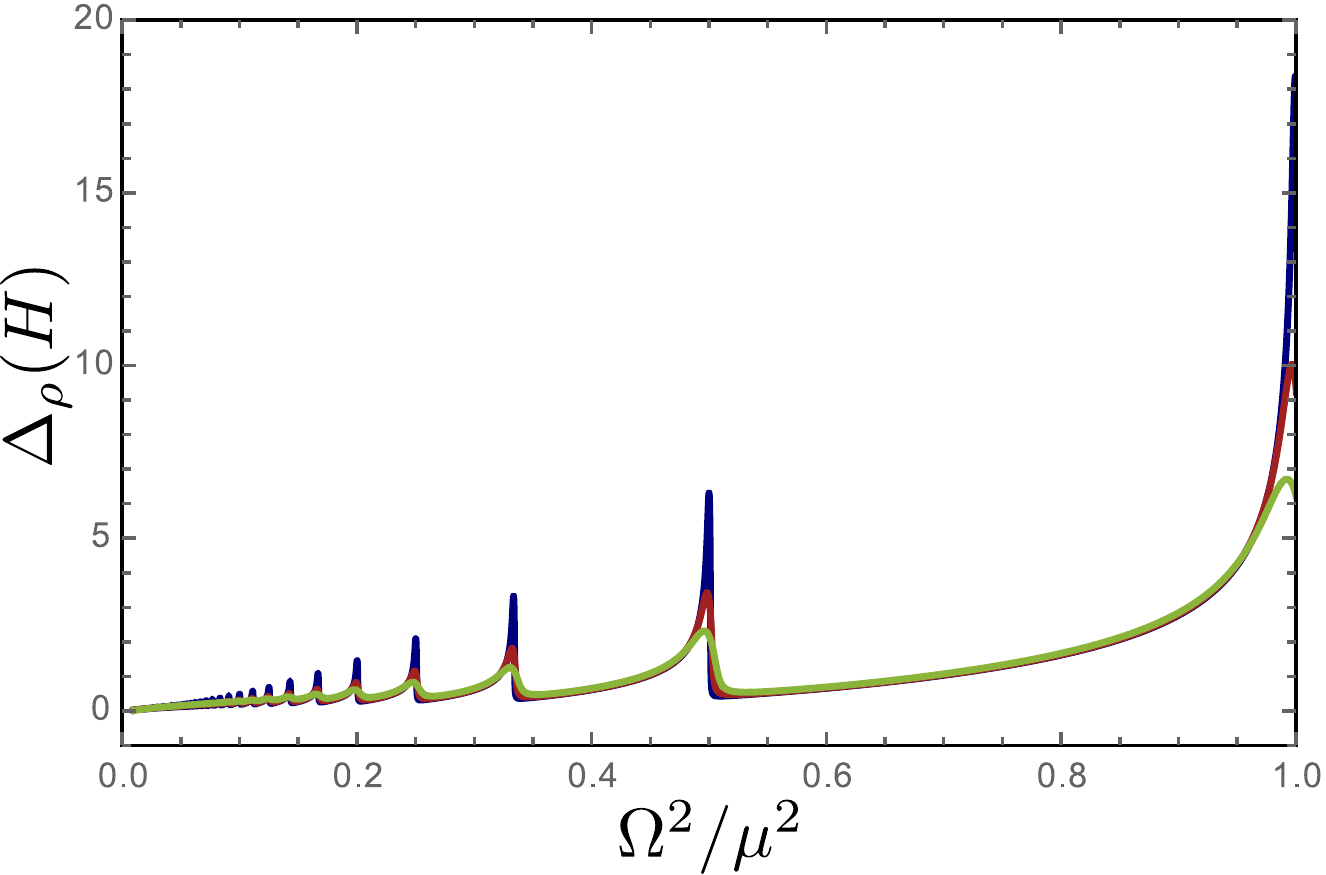}
\caption{Numerical evaluation of the TMR for pointlike impurities and fixed chemical potential in the regime of separated LLs. 
The figure depicts the numerical analysis of Eq.~\eqref{rhosep}. Blue, red, and green curves correspond to $A\mu/\Omega^2=5\cdot10^{-5},3\cdot10^{-4},1\cdot10^{-3}$, respectively. For all curves $\Lambda/\Omega^2=100$.}
\label{fig:magres-point-xy}
\end{figure}

We fix the values of $\mu$ and $\gamma$ and increase the magnetic field. A detailed evaluation of the TMR 
in different regimes is presented in Appendix~\ref{app:pointres} and summarized as follows:
\begin{align}\label{rho-short-range}
\Delta_\rho(H)\sim\left\lbrace\begin{array}{ll}
\dfrac{\Gamma^2(\mu)}{\gamma^2\mu^4}-1, &\quad \dfrac{\mu^{5/4}\gamma^{1/4}}{v^{3/4}}\ll\Omega<\mu,\\[0.4cm]
\dfrac{\Omega^2}{\mu^2}-1, &\quad \mu<\Omega\ll\dfrac{\mu^{1/2}v^{3/2}}{\gamma^{1/2}},\\[0.4cm]
\dfrac{v^6}{\gamma^2\Omega^2}, &\quad \dfrac{\mu^{1/2}v^{3/2}}{\gamma^{1/2}}\ll\Omega\ll\dfrac{v^3}{\gamma}.
\end{array}\right.
\end{align}
For $\mu>\Omega\gg\mu^{5/4}\gamma^{1/4}v^{-3/4}$, the function $\Gamma(\mu)$ is given by Eq.~\eqref{GammanAPP}, leading to the oscillations in the TMR from zero to a maximum value proportional to $H^{4/3}$. This behavior is visualized in Fig.~\ref{fig:magres-point-xy}.
For lower magnetic fields, $\Omega\ll\mu^{5/4}\gamma^{1/4}v^{-3/4}$, the TMR is given within the SCBA by an exponentially small correction, as discussed in Appendix~\ref{app:pointres}.

It is important to emphasize that the TMR is only large for the zeroth LL (for magnetic fields $\Omega>\mu$). For lower magnetic fields, a small background magnetoresistance is only present due to the different shape of the oscillations in the conductivity and the Hall conductivity and is zero for a smoothened curve. In the regime of the zeroth LL, the magnetoresistance first grows linearly with $H$ as long as the Hall conductivity is larger than $\sigma_{xx}$,
and then decays (being proportional to $H^{-1}$) in the limit of strongest $H$, where $\sigma_{xx}\gg\sigma_{xy}$. A schematic plot of the TMR is presented in Fig.~\ref{fig:magres-point-all}.

The effect of finite temperature for separated LLs is discussed in the end of Appendix~\ref{app:pointres}. There, we assume that temperature is still smaller than chemical potential, $T<\mu$, but larger than the distance between LLs, $T>\Omega/\sqrt{n}$ such that LLs are smeared by temperature. The magnetoresistance is small and linear in magnetic field for $\mu^{1/2}\gamma^{-1/2}\ll\Omega<\mu$, reading
\begin{align}\label{rho_finiteT}
\Delta_\rho(H)\sim\frac{\Omega^2}{\mu^2}.
\end{align}

\begin{figure}
\includegraphics[width=\linewidth]{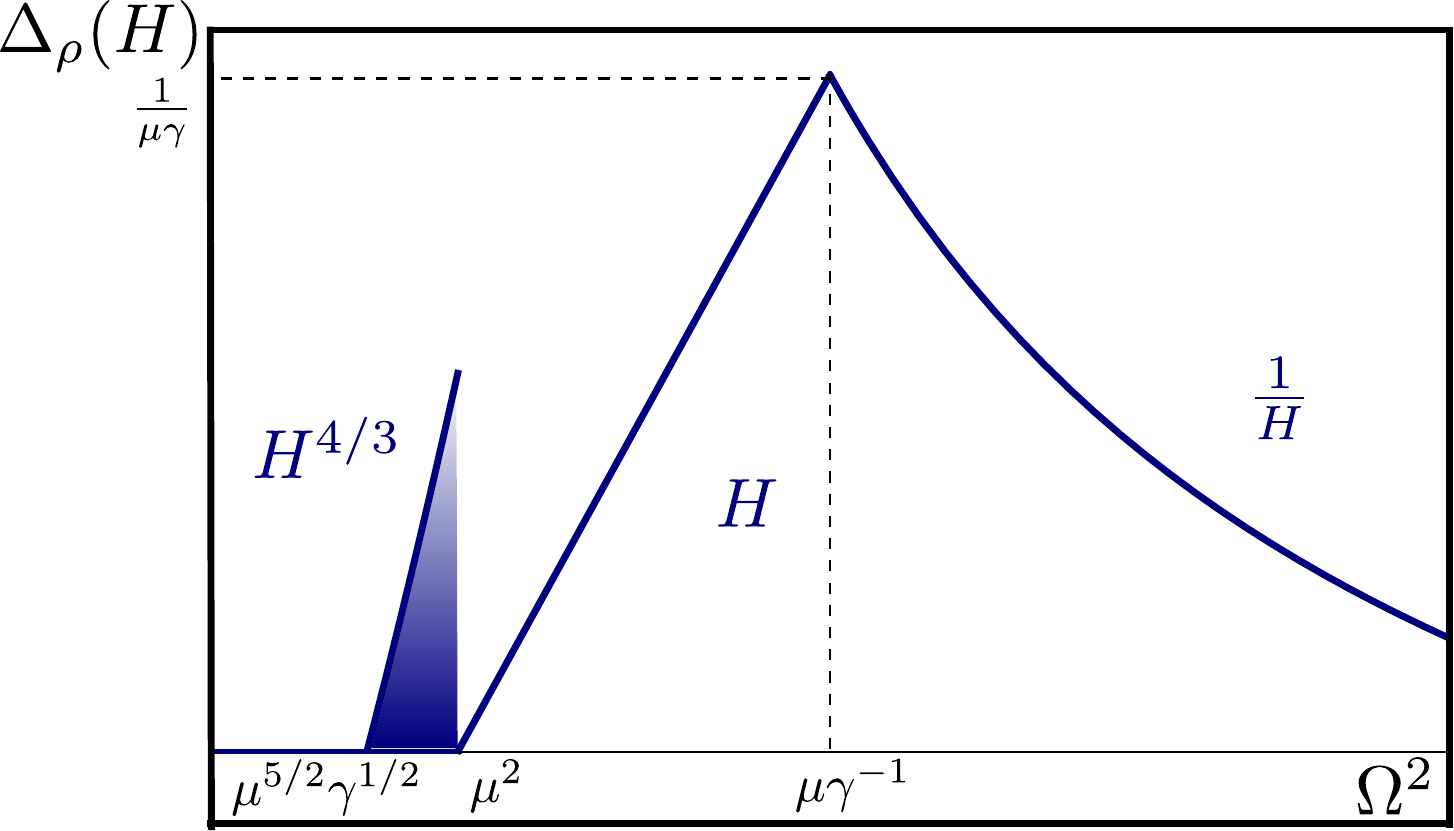}
\caption{Schematic illustration of the TMR for pointlike impurities and fixed chemical potential as given in Eq.~\eqref{rho-short-range}. The shaded region corresponds to oscillations of separated LLs described in Eq.~\eqref{rhopeak} and plotted in Fig.~\ref{fig:magres-point-xy}. }
\label{fig:magres-point-all}
\end{figure}

\begin{figure}
\includegraphics[width=\linewidth]{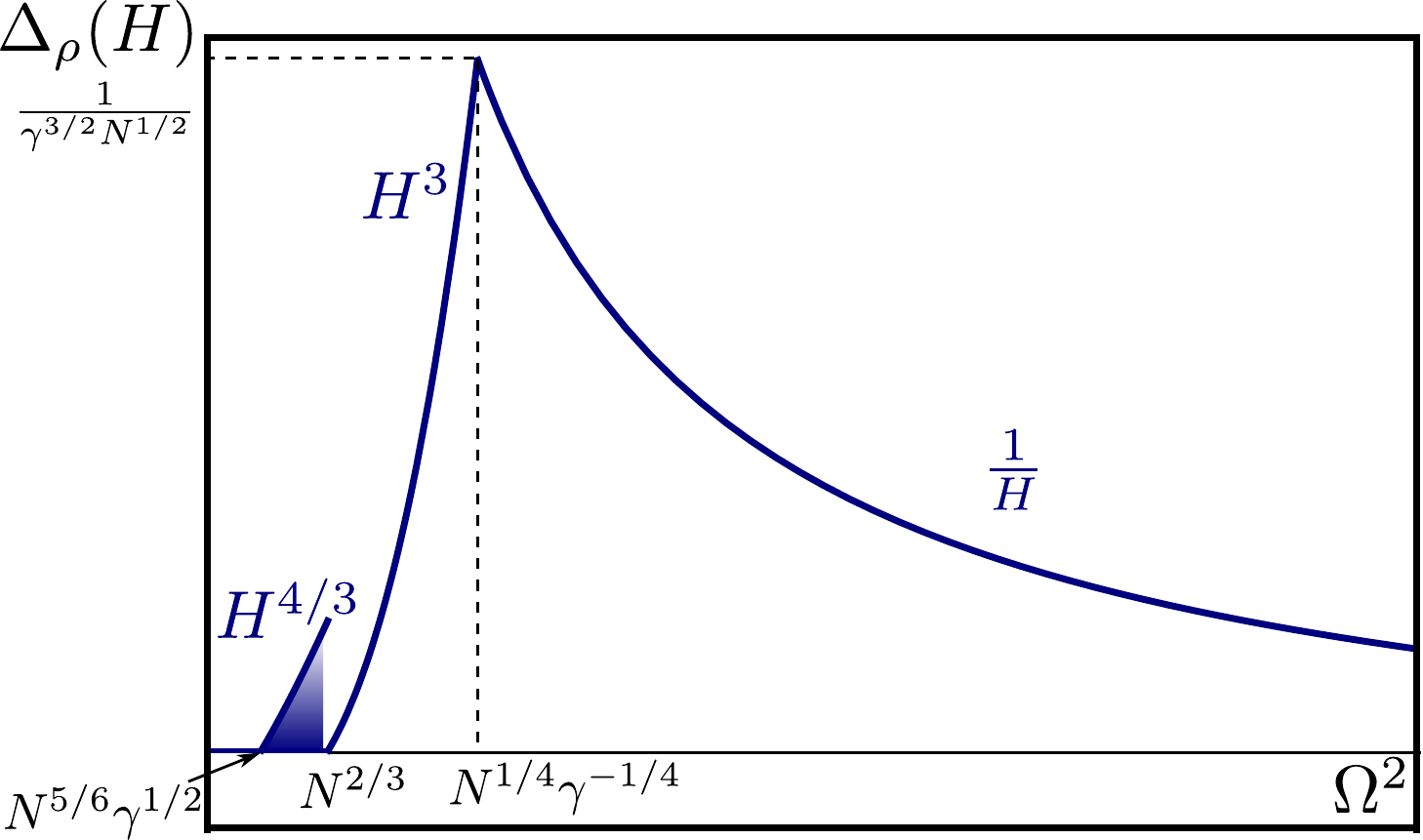}
\caption{Schematic illustration of the TMR for pointlike impurities and fixed particle density as given by Eq.~\eqref{rho-short-range-N}.
The shaded region corresponds to oscillations of separated LLs described by Eq.~\eqref{rhopeak-N}.}
\label{fig:magres-point-all-N}
\end{figure}

We continue now with an experimentally more relevant situation of a fixed particle density $N$. 
The details of the calculations are discussed in Appendix~\ref{app:pointres}; here we present the summary of results:
\begin{align}\label{rho-short-range-N}
\Delta_\rho(H)\sim\left\lbrace\begin{array}{ll}
\dfrac{\Gamma^2(N^{1/3})}{\gamma^2N^{4/3}}-1, &\ \dfrac{N^{5/12}\gamma^{1/4}}{v^{3/4}}\!\ll\Omega<\! N^{1/3},\\[0.4cm]
\dfrac{\Omega^6}{N^2}-1, &\ N^{1/3}<\!\Omega\ll \dfrac{N^{1/4}v^{3/4}}{\gamma^{1/4}},\\[0.4cm]
\dfrac{v^6}{\gamma^2\Omega^2}, &\ \dfrac{N^{1/4}v^{3/4}}{\gamma^{1/4}}\ll\Omega\ll\dfrac{v^3}{\gamma}.
\end{array}\right.
\end{align}
We observe that the behavior of the TMR at the fixed particle density only changes in the zeroth LL. 
For higher LLs, the particle density does not depend on magnetic field. 
The schematic behavior of the magnetoresistance is visualized in Fig.~\ref{fig:magres-point-all-N}

We conclude this section with a short discussion of the Hall resistivity $\rho_{xy}$ for fixed particle density, reading
\begin{align}\label{reshall}
\rho_{xy}=\frac{\sigma_{xy}}{\sigma_{xx}^2+\sigma_{xy}^2}.
\end{align}
For overlapping LLs, the anomalous Hall conductivity is exponentially small, see Eq.~\eqref{Nover}. 
In this regime, conductivity, Eq.~\eqref{conover}, and the normal Hall conductivity, Eq.~\eqref{Hallover} combine to
\begin{align}
\rho_{xy}\sim \frac{\Omega^2}{e^2 v^2 N}.
\label{rhoHall-linear}
\end{align}
For separated LLs the Hall conductivity, Eq.~\eqref{sigma-xy}, is larger than $\sigma_{xx}$ in magnetic fields up to $\Omega\sim N^{1/4}\gamma^{-1/4}$, 
again leading to Eq. (\ref{rhoHall-linear}).
For higher fields $\Omega>N^{1/4}\gamma^{-1/4}$, the conductivity of the lowest LL, Eq.~\eqref{con1}, has a large contribution resulting in
\begin{align}
\rho_{xy}\sim \frac{\gamma\Omega^6}{e^2 v^7 N^2}.
\end{align}
Therefore, the Hall resistivity shows a linear behavior up to the highest fields where it increases as a third power of magnetic field.

\section{Charged impurities}
\label{sec:charged}
\subsection{Screening}

In the previous parts of the paper we considered short-range impurities.
In this Section, we are going to generalize the obtained results for the case of screened Coulomb impurities that is expected to be particularly relevant experimentally.
The potential of a single Coulomb impurity is given by
\begin{equation}
U(\textbf{k})=\frac{4\pi e^{2}}{\epsilon_{\infty}(k^{2}+\kappa^{2})},
\end{equation}
where $\epsilon_{\infty}$ is the background dielectric constant and
$\kappa$ is the inverse screening radius that is determined by the thermodynamic density of states $\partial n/\partial \mu$ and reads, in the absence of disorder,
\begin{equation}
\label{kappa}
\kappa^{2}=\frac{4\pi e^2}{\epsilon_{\infty}}\frac{\partial n}{\partial\mu}
=\frac{e^{2}}{\pi\varepsilon_{\infty}v^3}
\left\{
  \begin{array}{ll}
    \Omega^2, &\quad \Omega\gg T,\mu, \\[0.2cm]
    \pi^2 T^2/3, &\quad T\ll \Omega,\mu, \\[0.2cm]
    \mu^2/2, &\quad \mu\ll\Omega,T.
  \end{array}
\right.
\end{equation}

In view of the singularity of Eq.~(\ref{kappa}) in the limit $\mu,T,H \to 0$, the effect of disorder becomes important, requiring a self-consistent treatment of disorder in the density of states.
The method is discussed in Ref.~\cite{PhysRevB.92.205113}, but, for the sake of clarity, we repeat the arguments below.

The analysis below is based on the assumption, that the ``fine-structure'' constant is not small,
\begin{equation}
e^2/(\varepsilon_\infty v)\agt 1.
\label{fine}
\end{equation}
In realistic situations with a fine-structure constant of the order of unity,
the characteristic values of $\kappa$ are of the order of $k_{\text{typical}}\sim \text{max}(\Omega, T)/v$, typical values
of the wave vector $k$.
With condition (\ref{fine}), the parametric dependence of the conductivity for the screened Coulomb disorder
is governed by an effectively pointlike correlator
 \begin{equation}\label{impcorr}
\left\langle U(\textbf{r})U(\textbf{r}')\right\rangle\simeq\gamma(H,T,\mu)\delta(\textbf{r}-\textbf{r}').
\end{equation}
From now on, we suppress the numerical prefactors.

The correlator (\ref{impcorr}) describes an effective white-noise disorder with the strength $\gamma(H,T,\mu)$
that depends on magnetic field, temperature, and chemical potential,
 \begin{align}\label{impcornew}
\gamma(H,T,\mu)&=N_{\text{imp}}
\left(\frac{\partial n}{\partial \mu}\right)^{-2}\notag\\
&\sim N_{\text{imp}}v^6
\left\{
  \begin{array}{ll}
    \Omega^{-4}, &\quad \Omega\gg T, \mu\\[0.2cm]
    T^{-4}, &\quad T\gg \Omega,\mu, \\[0.2cm]
    \mu^{-4} &\quad \mu\gg\Omega,T .
  \end{array}
\right.
\end{align}
Here $N_{\text{imp}}$ is the density of impurities.
In the limit $H,T,\mu\to 0$, Eq.\eqref{impcornew} leads to a divergent disorder strength.
Therefore, a self-consistent treatment of the impurity screening becomes necessary. At
\begin{align}
\text{max}(\Omega,T,\mu)\sim\varepsilon_\text{imp}=N^{1/3}_\text{imp}v,
\end{align}
the impurity-induced density of state will determine the screening
\begin{align}
\gamma(H,T,\mu)\sim\frac{v^3}{\varepsilon_\text{imp}}.
\end{align}
For Coulomb impurities, the weak disorder regime is valid under the condition $\text{max}(\Omega, T,\mu)\gtrsim\varepsilon_\text{imp}$.
Under these conditions, the results of the previous sections
 are applicable to the Coulomb case with the replacement of $\gamma$ with $\gamma(H,T,\mu)$.
The dependence of the strength of screened disorder on magnetic filed plays a crucial role in the $H$ dependence of TMR
for charged impurities.

\subsection{Magnetoresistance}
In order to find the TMR, we substitute $\gamma(H,\mu)\sim\varepsilon_\text{imp}^3v^3[\text{max}(\Omega,\mu)]^{-4}$ for $\gamma$ in the conductivity and Hall conductivity. As we do not keep numerical prefactors, the vertex corrections can be disregarded (since they only modify these prefactors). The particular substitution in each regime is done in Appendix~\ref{App:charged-xx-xy}; here we only state the results.

For charged impurities, we need to distinguish between $\mu>\varepsilon_\text{imp}$ and $\mu<\varepsilon_\text{imp}$.
We start with the case of $\mu<\varepsilon_\text{imp}$, where only overlapping LLs and the zeroth LLs are important. 
We fix $\mu$ and $\varepsilon_\text{imp}$ while increasing the magnetic field.  The TMR reads
 \begin{align}\label{magreslow}
\Delta_\rho(H)&\sim \left\{
  \begin{array}{ll}
\dfrac{\varepsilon_\text{imp}^4}{\mu^2\Omega^2}, &\quad  \dfrac{\varepsilon_\text{imp}^{3/2}}{\mu^{1/2}}\ll\Omega\\[0.3cm]
    \dfrac{\Omega^2}{\varepsilon_\text{imp}^2}, &\quad \varepsilon_\text{imp}\ll\Omega\ll\dfrac{\varepsilon_\text{imp}^{3/2}}{\mu^{1/2}}.
  \end{array}
\right.
\end{align}
The TMR for lower magnetic fields is exponentially small for the same reason as for pointlike impurities. 
We find a linear TMR for the zeroth LL and Coulomb impurities in fields up to $\varepsilon_\text{imp}^{3/2}\mu^{1/2}$.
In the highest magnetic fields, the TMR at fixed chemical potential vanishes as $H^{-1}$. We will see below that the behavior of the
TMR in the ultra-quantum limit is different for the case of a fixed density, where the TMR keeps growing linearly.  

\begin{figure}
\begin{center}
\includegraphics[width=\linewidth]{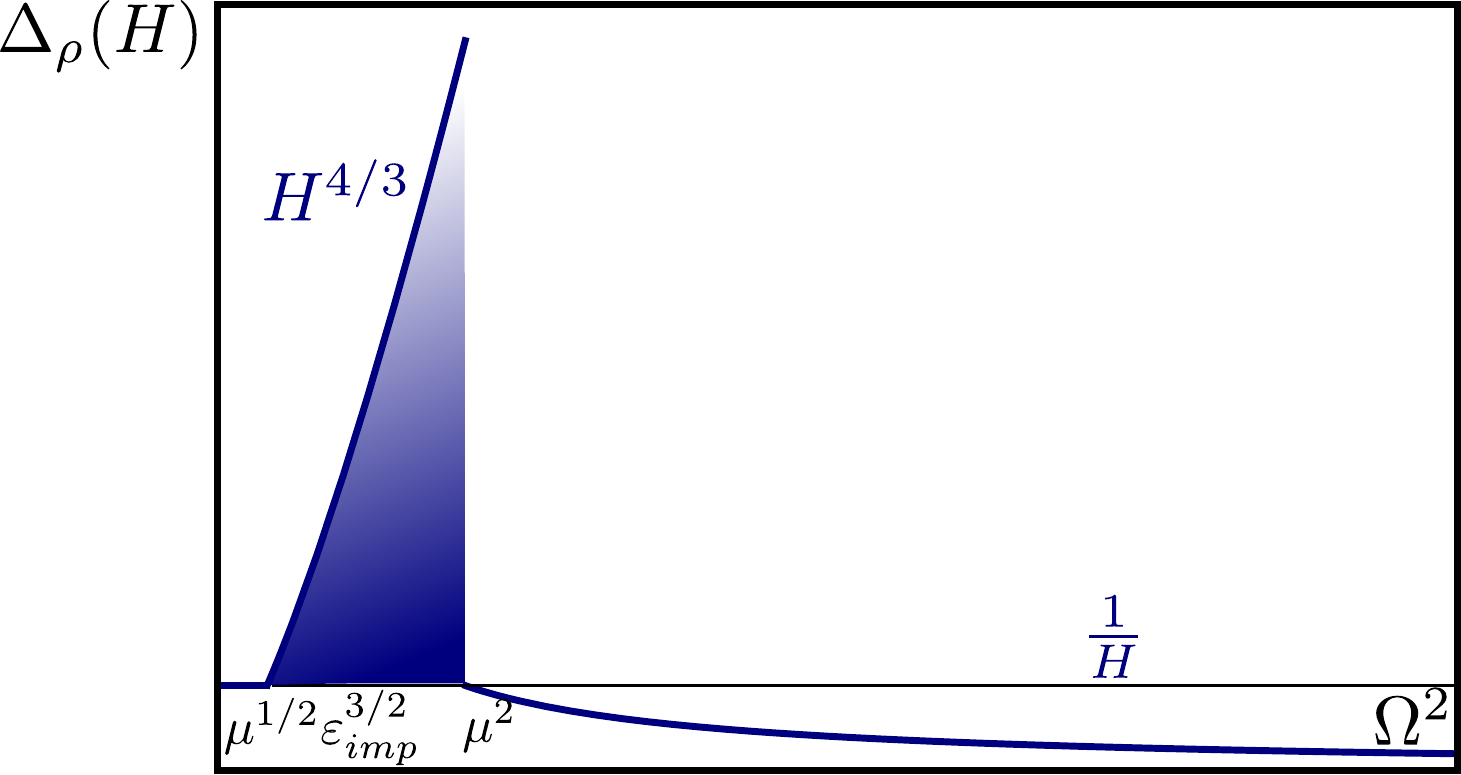}
\caption{Schematic illustration of TMR for Coulomb impurities at a fixed chemical potential, $\mu\gg \varepsilon_\text{imp}$.
At lowest fields, the TMR is determined by separated LLs with large background.
With increasing magnetic field, first separated LLs give rise to the peaks in the TMR (indicated by the shaded region), and finally the zeroth LL becomes relevant for transport. The scaling of the TMR in various regimes is given by Eq.~\eqref{magreshigh}. }
\label{fig:resmu}
\end{center}
\end{figure}

In the opposite regime, $\mu>\varepsilon_\text{imp}$, we have both regimes of separated LLs and separated LLs with 
the dominating background density of states. The TMR reads
\begin{align}\label{magreshigh}
\Delta_\rho(H)\sim\left\lbrace\begin{array}{ll}
\dfrac{\Gamma_C^2\mu^4}{\varepsilon_\text{imp}^6}-1, &\quad \mu^{1/4}\varepsilon_\text{imp}^{3/4}\ll\Omega<\mu,\\[0.4cm]
\dfrac{\mu^2}{\Omega^2}-1, &\quad \mu<\Omega,
\end{array}\right.
\end{align}
where $\Gamma_C(\mu,\Omega)$ defines the oscillations of the conductivity and is defined in Eq.~\eqref{SCBAcoul}.
As for pointlike impurities, the TMR is vanishing small for lower magnetic fields. Furthermore, the magnetic field dependence of the TMR changes only for the lowest LL, $\Omega>\mu$ compared to pointlike impurities. For the lowest LL, the screening is magnetic-field dependent, while for higher LLs, the screening is dominated by chemical potential. Therefore, Fig.~\ref{fig:magres-point-xy} can be redrawn just by changing the dependence on chemical potential and disorder strength $\varepsilon_\text{imp}$. A schematic plot of the TMR is presented in Fig.~\ref{fig:resmu}.

\begin{figure}
\begin{center}
\includegraphics[width=\linewidth]{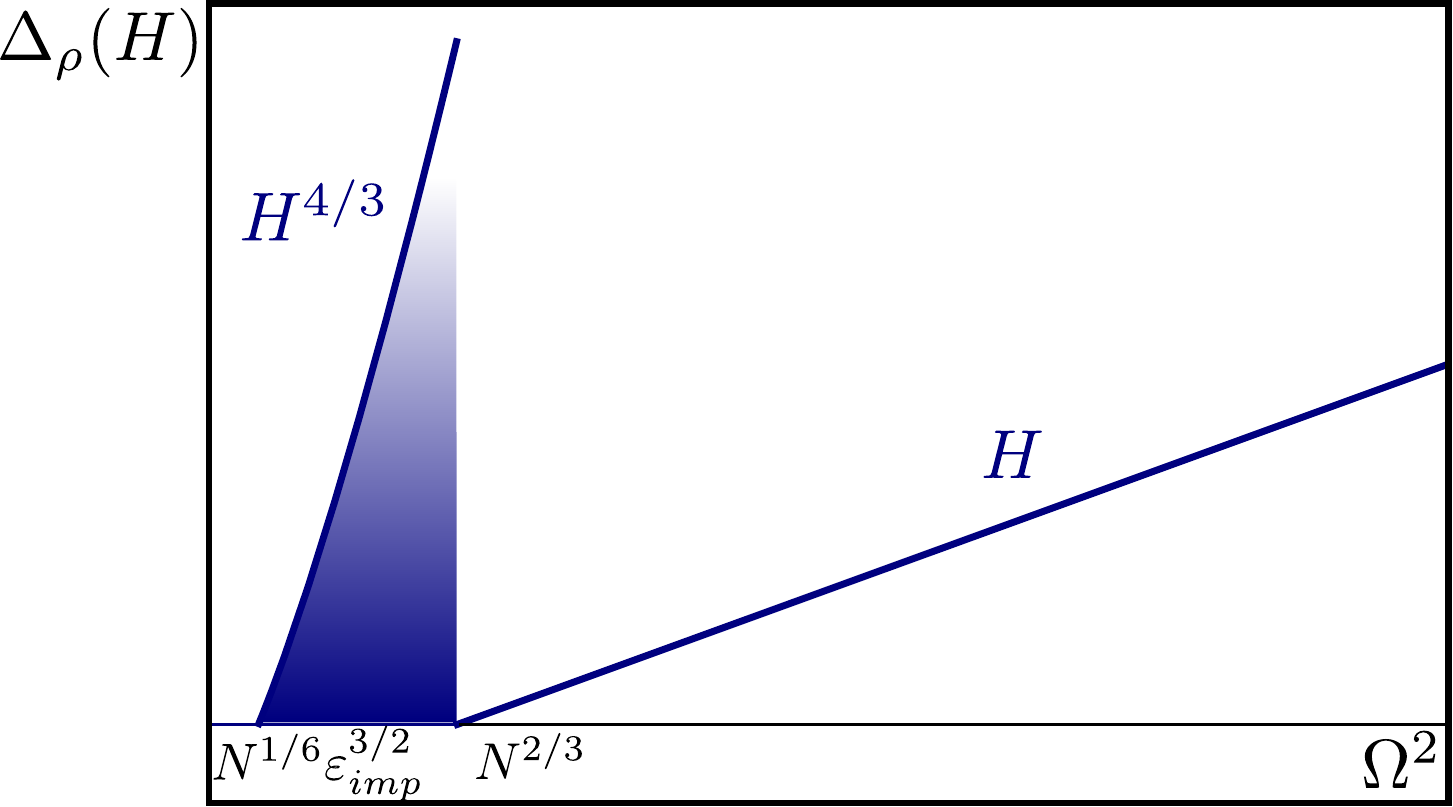}
\caption{Schematic illustration of TMR for Coulomb impurities at a fixed particle density, $N\gg \varepsilon_\text{imp}^3$.
The corresponding scaling is given by Eq.~\eqref{magreshigh-N}.}
\label{resN}
\end{center}
\end{figure}

\begin{figure*}
\begin{center}
\includegraphics[width=8.5cm]{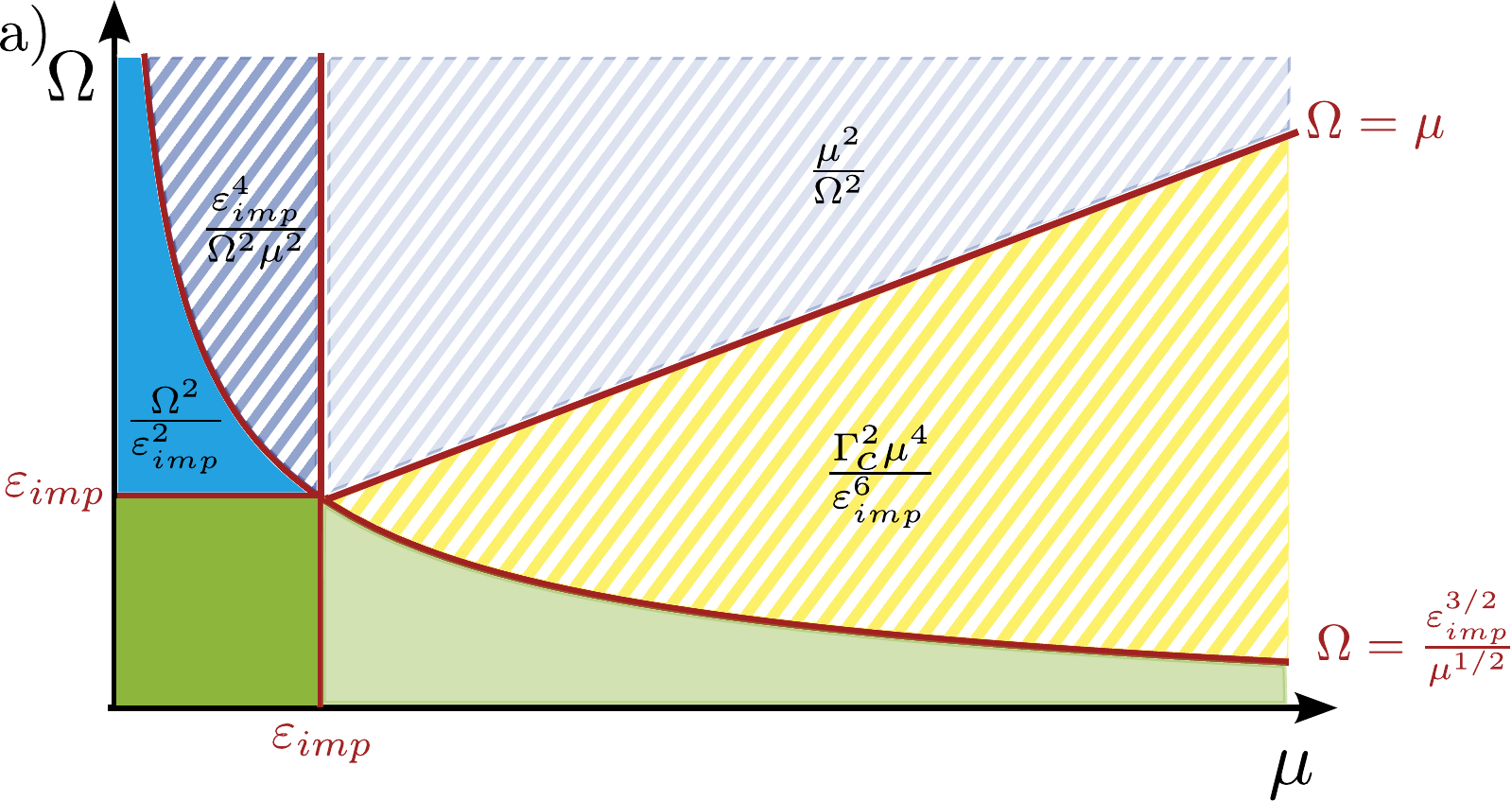}
\includegraphics[width=8.5cm]{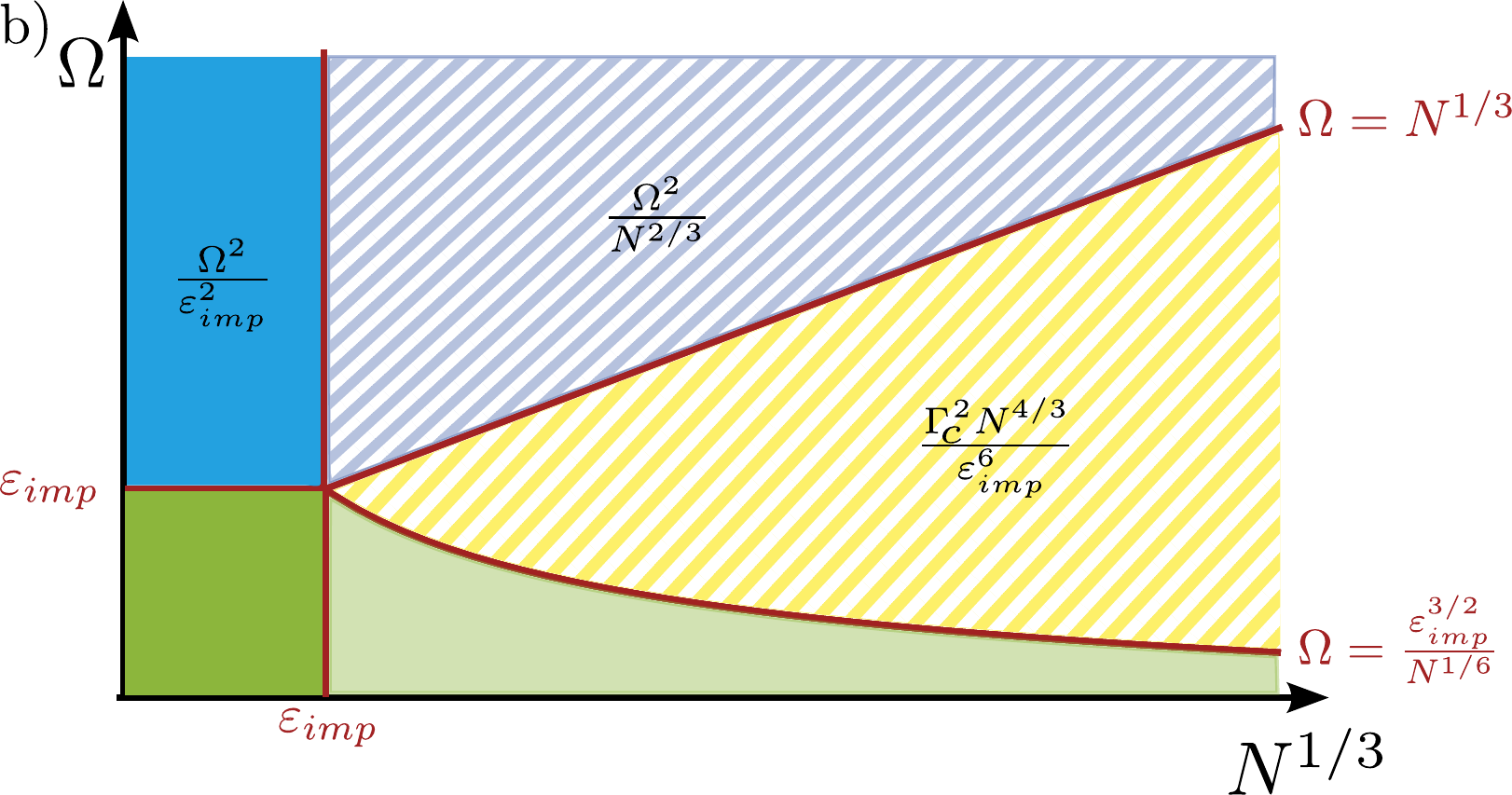}
\caption{Behavior of the TMR $\Delta_\rho(H)$ in a Weyl semimetal with Coulomb impurities for fixed chemical potential (a) and for fixed particle density (b). Scaling of dominant contribution to the TMR in each of the parameter regimes (the Fermi velocity $v$ is set to unity) and
equations for borderlines between the regimes are indicated. The striped (filled) parts indicate the regions where $\sigma_{xy}$ ($\sigma_{xx}$) dominates
the denominator in Eq.~(\ref{magres}) for the TMR $\Delta_\rho(H)$. The blue regions correspond to the regime of zeroth LL, yellow -- separated LLs, and green -- overlapping LLs. The SdHO in the yellow region are due to the magnetooscillations of the scattering rate $\Gamma_C$ defined in Eq.~\eqref{SCBAcoul}.}
\label{phasediagramCoul}
\end{center}
\end{figure*}

If we fix the particle density as relevant for experiments, the magnetic-field dependence of the resistivity of the zeroth LL 
changes because of magnetic-field dependence of the particle density. The TMR for $N^{1/3}<\varepsilon_\text{imp}$ reads
\begin{align}\label{magreslow-N}
\Delta_\rho(H)&\sim
  \begin{array}{ll}
    \dfrac{\Omega^2}{\varepsilon_\text{imp}^2}, &\quad \Omega\gg\varepsilon_\text{imp}.
    \end{array}
\end{align}
The TMR in the limit of highest magnetic fields is linear in $H$, which agrees with the results of Refs.~\cite{PhysRevB.58.2788,PhysRevB.92.205113}.

For $N^{1/3}>\varepsilon_\text{imp}$, the TMR is given by
\begin{align}\label{magreshigh-N}
\Delta_\rho(H)\sim\left\lbrace\begin{array}{ll}
\dfrac{\Gamma_C^2 N^{4/3}}{\varepsilon_\text{imp}^6}-1, &\ N^{1/12}\varepsilon_\text{imp}^{3/4}\ll\Omega<N^{1/3},\\[0.4cm]
\dfrac{\Omega^2}{N^{2/3}}-1, &\ N^{1/3}<\Omega.
\end{array}\right.
\end{align}
The resulting linear TMR in highest magnetic fields is in agreement with Eq.~\eqref{magreslow-N} and with Refs.~\cite{PhysRevB.58.2788,2016arXiv160704943X,PhysRevB.92.205113}. 
The only difference is the replacement of the disorder scale $\varepsilon_\text{imp}$ in the slope of the TMR with $N^{1/3}$.
The TMR in the lower fields remains vanishing with small oscillations, see Fig.~\ref{resN}.
The resulting ``phase diagrams'' for TMR in the cases of fixed chemical potential and fixed density are presented in Figs. \ref{phasediagramCoul}a and \ref{phasediagramCoul}b, respectively.

For finite temperature under the conditions $T<\mu$ and $T>\Omega/\sqrt{n}$, the magnetoresistance calculated in Appendix~\ref{app:pointres} applies here. For fixed particle density, the magnetoresistance given by Eq.~\eqref{rho_finiteT} for $N^{1/12}\varepsilon_\text{imp}^{3/4}\ll\Omega<N^{1/3}$ scales in the same way as the linear magnetoresistance for the zeroth LL [cf. second line of Eq.~\eqref{magreshigh-N}]:
\begin{align}\label{rho_finiteTcoul}
\Delta_\rho(H)\sim\frac{\Omega^2}{N^{2/3}}.
\end{align}

Finally, we address the Hall resistivity, Eq.~\eqref{reshall}, for fixed particle density. 
Similarly to the case of pointlike impurities, the conductivity and Hall conductivity away from the quantum limit combine into
\begin{align}
\rho_{xy}\sim\frac{\Omega^2}{e^2v^2 N}.
\end{align}
In the quantum limit, $\Omega>N^{1/3}$,  conductivity, Eq.~\eqref{concoul1}, and Hall conductivity, Eq.~\eqref{sigma-xy}, scale identically with magnetic field. Therefore, the Hall resistivity  is
\begin{align}
\rho_{xy}\sim\frac{N\Omega^2}{e^2v^2\varepsilon_\text{imp}^4}
\end{align}
for $\varepsilon_\text{imp}>N^{1/3}$
and
\begin{align}
\rho_{xy}\sim\frac{\Omega^2}{e^2v^2 N}
\end{align}
for $\varepsilon_\text{imp}<N^{1/3}$.
In the physically most relevant situation where a finite particle density is induced by donors (charged impurities), $\varepsilon_\text{imp}\sim N^{1/3}$,
the Hall resistivity is of the same order as the TMR.

To conclude this section, we outline its main findings:
\begin{itemize}
\item[(i)] For Coulomb impurities, the TMR is linear in the ultra-quantum limit;
\item[(ii)] In the experimentally relevant case, $\varepsilon_\text{imp}\sim N^{1/3}$, the Hall resistivity is of the same order as the TMR;
\item[(iii)] Strong SdHO are observed in moderate magnetic fields, where the background TMR is negligible.
\end{itemize}
All these findings are in agreement with the numerical results of Ref.~\cite{2016arXiv160704943X}.
The results (i) and (ii) conform with the experimental observations \cite{RIS_5} of a strong linear TMR comparable to the Hall resistivity. 
However, the above model treated within the SCBA does not explain the emergence of the
SdHO on top of rapidly growing background TMR as observed in experiments, contrary to (iii).
In the next section, we propose a model that can explain such a behavior.

\section{Magnetoresistance for shifted Weyl nodes}
\label{sec:shifted}

We discuss now a model with Weyl nodes shifted in energy, see Fig.~\ref{energybands}. 
In various experiments \cite{RIS_5, 2016arXiv161001413C}, the different pairs of Weyl nodes are shifted in energy with respect to each other 
such that some pairs of nodes are characterized by a positive chemical potential, whereas other nodes by a negative chemical potential 
counted from the corresponding nodal points.

The conductivity $\sigma_{xx}$ is an even function of magnetic field and does not depend on the sign of the chemical potential in a particle-hole symmetric spectrum, so that the contributions of different nodes to $\sigma_{xx}$ just add up. Even exactly at charge neutrality, the conductivity of each pair of nodes
 is determined by a finite density of quasiparticles (electrons or holes, $N_+$ and $N_-$, respectively), similarly to the consideration of a single node above. It is important to notice that away from charge neutrality the SdHO show a superposition of oscillations from the pairs of nodes characterized by the different chemical potentials.
At the same time, the Hall conductivity is an odd function of chemical potential and hence vanishes at charge neutrality. 
Therefore, the distance to the complete charge compensation point, which is in realistic cases typically smaller than the chemical potential of each pair of nodes (see discussion in Ref.~\cite{RIS_5}), is of crucial importance for the Hall response. 

We will first discuss the case, when the chemical potentials of the different nodes correspond to the charge compensation point characterized by 
a vanishing Hall conductivity, $\sigma_{xy}=0$.
The magnetoresistance is then fully determined by the conductivity $\sigma_{xx}$,
\begin{align}
\Delta_\rho(H)=\frac{\sigma_{xx}(0)}{\sigma_{xx}(H)}-1
\end{align}
As we have assumed that the carriers in one pair of Weyl node has the chemical potential $\Delta$ while in the other pair the chemical potential is $-\Delta$, as depicted in Fig.~\ref{energybands}, the total TMR is multiplied by the number of Weyl nodes.

\subsection{Pointlike impurities}

\begin{figure}
\begin{center}
\includegraphics[width=8cm]{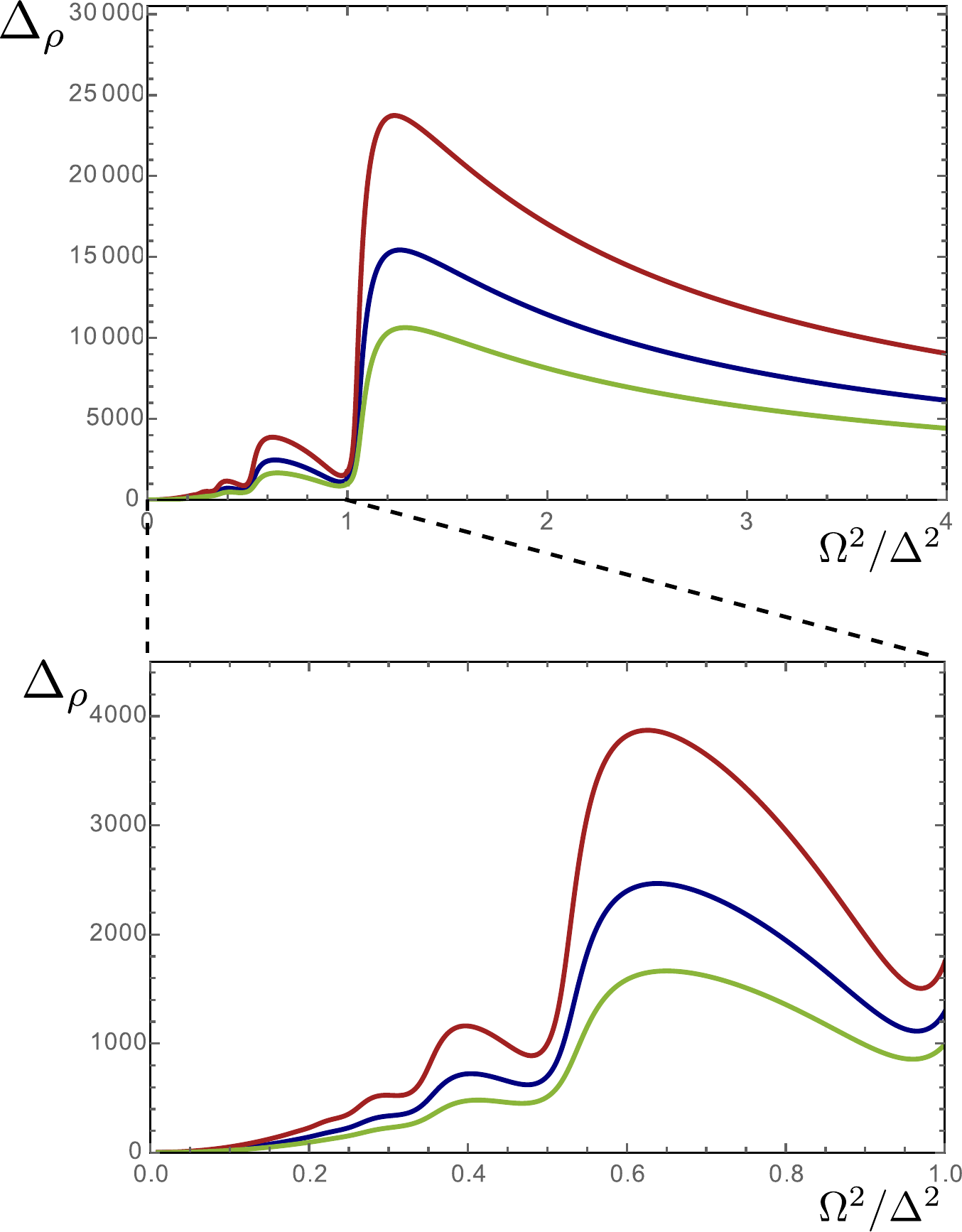}
\end{center}
\caption{TMR for separated LLs as a function of $\Omega^2/\Delta^2$ for pointlike impurities and for Weyl nodes shifted in energy by $2\Delta$.
The results are obtained by using Eq.~\eqref{con1} for $\Delta<\Omega$ and Eq.~\eqref{consep} for $\Delta>\Omega$.
Red, blue, and green lines correspond to $A\Delta/\Omega^2=5\cdot10^{-3},6\cdot10^{-3},7\cdot10^{-3}$, respectively.
For all curves $\Lambda/\Omega^2=100$.}
\label{magnressep}
\end{figure}

To obtain the TMR for the case of zero Hall conductivity, we use Eq.~\eqref{consum} for the conductivity in the different regimes.
We fix now the values of $\Delta$ and $\gamma$ and analyze the evolution of the TMR with increasing magnetic field:
\begin{equation}
\Delta_\rho\sim
\left\{
  \begin{array}{ll}
       \!\dfrac{v^6\Omega^4}{\Delta^6\gamma^2}-1, &\   \dfrac{\gamma^{1/2}\Delta^{3/2}}{v^{3/2}}\!\ll\Omega\ll\!\dfrac{\gamma^{1/4}\Delta^{5/4}}{v^{3/4}}, \\[0.3cm]
   \!\dfrac{\Omega^4}{\Delta^2\Gamma^2(\Delta)}-1, &\  \dfrac{\gamma^{1/4}\Delta^{5/4}}{v^{3/4}}\ll \Omega < \Delta, \\[0.3cm]
   \!\dfrac{v^6}{\gamma^2\Omega^2}, &\   \Delta < \Omega \ll \dfrac{v^3}{\gamma} .
  \end{array}
\right.
\label{short-range-deltarho}
\end{equation}
In the regime of separated LLs, $\gamma^{1/4}\Delta^{5/4}v^{-3/4}\ll \Omega \ll \Delta$, we find a sublinear ($H^{2/3}$) behavior of the minima of the SdHO in the TMR, while the maxima of the TMR show a quadratic growth with magnetic field.
The SdHO and the TMR as obtained from the numerical solution of the SCBA equations are depicted in Fig.~\ref{magnressep}. 
In the limit of highest magnetic field, the TMR decays as $1/H$, similarly to the case of non-shifted Weyl nodes.

\subsection{Charged impurities}

The condition of overall charge neutrality of the sample at zero charge of carriers ($N_+=N_-$) can be maintained for a finite
concentration of Coulomb impurities when the concentration of positively and negatively charged impurities are equal.
The conductivity for Coulomb impurities is analyzed in Appendix~\ref{App:charged-xx-xy} and is given by Eqs.~\eqref{sigma-xx-coul-low} and \eqref{sigma-xx-coul-high}. For fixed values of $\Delta$ and $\varepsilon_\text{imp}$, we first calculate the TMR 
 for $\Delta<\varepsilon_\text{imp}$:
\begin{align}
\Delta_\rho\sim
  \begin{array}{ll}
   \dfrac{\Omega^2}{\varepsilon_\text{imp}^2}, &\   \varepsilon_\text{imp} \ll \Omega.
  \end{array}
\end{align}
For $\Delta>\varepsilon_\text{imp}$, the evolution of the TMR with increasing magnetic field is described by
\begin{equation}
\Delta_\rho\sim
\left\{
  \begin{array}{ll}
       \dfrac{\Omega^4\Delta^2}{\varepsilon_\text{imp}^6}-1, &\   \dfrac{\varepsilon_\text{imp}^{3/2}}{\Delta^{1/2}}\ll \Omega \ll \Delta^{1/4}\varepsilon_\text{imp}^{3/4}, \\[0.3cm]
   \dfrac{\Omega^4}{\Delta^2\Gamma_C^2(\Delta,\Omega)}-1, &\   \Delta^{1/4}\varepsilon_\text{imp}^{3/4}\ll \Omega < \Delta, \\[0.3cm]
   \dfrac{\Omega^2\Delta^4}{\varepsilon_\text{imp}^6}, &\   \Delta < \Omega .
  \end{array}
\right.
\label{charged-deltarho}
\end{equation}
In both limits we find a large, linear TMR in the quantum limit where only the lowest LL contributes to transport. 
We observe that the linear TMR in highest magnetic fields is very robust and does not depend on whether the Weyl nodes are shifted 
in energy or not (cf. Sec.~\ref{sec:charged}).

\begin{figure}
\begin{center}
\includegraphics[width=\linewidth]{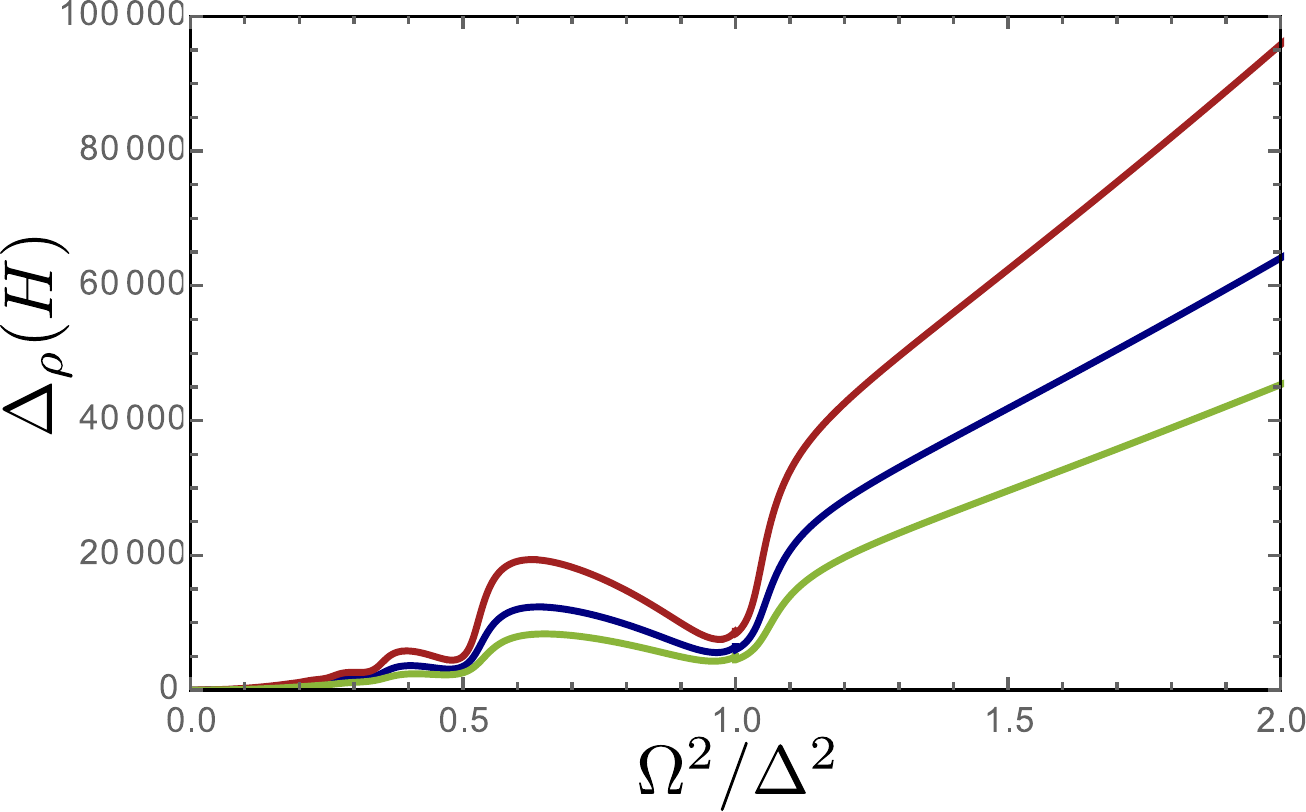}
\caption{TMR for Coulomb impurities and shifted Weyl nodes as a function of $\Omega^2/\Delta^2$.
The results are obtained from Eq.~\eqref{concoul1} for $\Omega>\Delta$ and from Eq.~\eqref{concoulsep} for $\Omega<\Delta$.
These results nicely match at the border of the regimes in the numerical evaluation.
Red, blue and green lines correspond to $\varepsilon_\text{imp}^3/\Delta^3=5\cdot10^{-3},6\cdot10^{-3},7\cdot10^{-3}$, respectively.
For all curves $\Lambda/\Omega^2=100$.}
\label{fig:magnrescoul}
\end{center}
\end{figure}

For lower magnetic field, the result is similar to the case of pointlike impurities. The minima of the TMR 
evolve as $H^{2/3}$ and the maxima as $H^2$ in magnetic field.
The result of numerical evaluation of TMR is depicted in Fig.~\ref{fig:magnrescoul} showing both the 
magnetooscillations and the TMR in the ultra-quantum limit. The overall picture of the TMR agrees with the behavior found in experiments.
Specifically, with increasing magnetic field, the TMR shows strong SdHO on top of the rapidly growing background and crosses over into a purely linear TMR
without magnetooscillations in the limit of highest magnetic field.

Away from the exact compensation point, where the Hall resistivity is finite,
the above picture for the TMR with SdHO on top of strong TMR remains intact as long as $\sigma_{xx}\gg \sigma_{xy}$. 
Denoting by $\delta\mu\propto (N_+-N_-)^{1/3}\ll \Delta$ the distance from the neutrality point, we get
\begin{equation}\label{sigma_xy_mu_shifted}
\sigma_{xy}\sim \frac{e^2 \delta\mu}{v}\left\{
\begin{array}{ll}
1, & \Delta<\Omega, \\[0.2cm]
\dfrac{\Delta^2}{\Omega^2}, & \Delta > \Omega.\\[0.2cm]
\end{array}\right.
\end{equation}
The condition $\sigma_{xx}\gg \sigma_{xy}$ translates with the background 
conductivity given by Eq.~(\ref{concoulpeak}) at $\Delta>\Omega$ into
\begin{equation}
\frac{\varepsilon_\text{imp}^3}{\Omega^2}\gg \delta \mu.
\end{equation}
This can be fulfilled in a broad range of magnetic fields when the concentrations of positively and negatively 
charged impurities are close.

\section{Summary and Discussion}
\label{sec:summary}

To summarize, we have generalized the theory of the transverse magnetoresistivity of Weyl semimetals developed in Ref.~\cite{PhysRevB.92.205113} to the case of a finite chemical potential (finite carrier density). We have considered two models of disorder: (i) short-range impurities and (ii) charged (Coulomb) impurities. Away from charge neutrality, the analysis includes the calculation of the Hall conductivity and the Shubnikov-de Haas oscillations.
We have further extended the consideration to a realistic model with Weyl nodes shifted in energy (as found in various Dirac and Weyl materials)
with the chemical potential corresponding to the total charge neutrality.
We have identified a rich variety of regimes of the resistivity scaling in the plane spanned by the magnetic field and the chemical potential (or carrier density) that emerge because of the unusual broadening of Landau levels and are governed by a competition between the conductivity $\sigma_{xx}$ and Hall conductivity $\sigma_{xy}$. We have also found that the TMR in strongest magnetic fields depends on whether the particle density or the chemical potential is fixed.

For pointlike impurities, the TMR is negligible in moderate magnetic fields (even for separated Landau levels), showing 
peaks at the centers of Landau levels. 
A pronounced magnetoresistance is only observed for the zeroth Landau level, where the TMR decays as $1/H$ in the ultra-quantum limit 
for both fixed chemical potential and fixed particle density, see Figs.~\ref{fig:magres-point-xy} and \ref{fig:magres-point-all}.

In the model of Coulomb impurities (which is expected to be more relevant experimentally), while the behavior of the TMR in moderate magnetic fields is similar,
the crucial difference appears in strongest magnetic fields, where a linear-in-$H$ TMR emerges, see Figs.~\ref{fig:resmu}-\ref{phasediagramCoul}. For a fixed chemical potential, the TMR is linear only in a finite range of $H$ in a close vicinity of the charge neutrality point and decreases with magnetic field as $1/H$ otherwise.
For a fixed particle density (which should be the case in experiments), we obtain in the ultra-quantum limit a nonsaturating linear TMR of the type first discovered in Ref.~\cite{PhysRevB.58.2788}. While the prefactor of  the linear TMR away form the neutrality point is different from that at charge neutrality, Ref.~\cite{PhysRevB.92.205113}, the scaling with magnetic field is the same. Moreover, the conductivity and Hall conductivity are of the same order in the experimentally relevant situation \cite{RIS_5, PhysRevB.91.041203} of particle density being roughly equal to the concentration of impurities,
$N\sim\varepsilon_\text{imp}^3$. Within this model, the range of magnetic fields where the Shubnikov-de Haas oscillation are developed corresponds to a
weak background TMR, while the strong (linear) TMR emerges only in the ultra-quantum limit, where only the zeroth Landau level contributes to transport and
hence no magnetooscillations can be observed.

Further, we have analyzed a more sophisticated (but experimentally relevant) model which describes different pairs of Weyl nodes 
shifted in energy with respect to each other (Fig.~\ref{energybands}).
In such systems, the Hall conductivity can be partly or fully compensated, 
while $\sigma_{xx}$ in each pair of nodes corresponds to a finite density of quasiparticles.
Within this model, the range of moderate magnetic fields, where Shubnikov-de Haas oscillations become strong, overlaps with the range of fields where the background TMR grows rapidly. The minima of the oscillations evolve with magnetic field as $H^{2/3}$ while the maxima increase quadratically with magnetic field. This holds for both models of disorder (pointlike and Coulomb impurities). Thus for shifted pairs of Weyl nodes,
an intermediate regime of magnetic fields emerges where the Shubnikov - de Haas oscillations are superimposed on the strong background magnetoresistance originating from separated Landau levels. (This is impossible in the case of ultra-quantum linear TMR of Refs.~\cite{PhysRevB.58.2788,PhysRevB.92.205113} which is entirely governed by the lowest LL.) The difference between the two models of disorder manifests itself in the 
ultra-quantum limit where only the zeroth LL contributes to transport. There, we find a decay of the magnetoresistance proportional to $1/H$ for pointlike impurities and a large, linear magnetoresistance for charged impurities, consistent with those found in Refs.~\cite{PhysRevB.58.2788,PhysRevB.92.205113}.  The results for TMR in the two different models of disorder are visualized in Figs.~\ref{magnressep} and \ref{fig:magnrescoul}.
We emphasize that this work focused on the two idealized cases: (i) no charge compensation between different pairs of Weyl nodes, (ii) complete charge compensation between the different pairs of nodes. The intermediate case of a partial compensations as present in experiments \cite{RIS_5,2016arXiv161001413C} would show a variety of effects governed by the competition between $\sigma_{xx}$ and $\sigma_{xy}$ as well as a superposition of Shubnikov-de Haas oscillations
coming from different nodes. 

It should be noted that the calculations in this paper have been mainly performed at zero temperature. As usual, finite temperature smears Shubnikov-de Haas oscillations. Our results are well applicable for temperatures smaller than the distance between neighboring Landau levels -- in the regime where pronounced Shubnikov-de Haas oscillations are observed. There the finite temperature only leads to a small correction while keeping the background TMR essentially unchanged. 
We also briefly discussed the effect of thermal smearing at higher temperatures where the thermal averaging exponentially suppresses the magnetooscillations, and leads to a finite background TMR
even in the model of non-shifted Weyl nodes, similarly to the case of charge neutrality \cite{PhysRevB.92.205113}. The TMR is small and linear in the regime of thermal averaging [cf. Eqs.~\eqref{rho_finiteT} and \eqref{rho_finiteTcoul}]. For Coulomb impurities, this thermally smeared quantum TMR, Eq.~\eqref{rho_finiteTcoul} scales in the same way as the ultra-quantum TMR, second line of Eq.~\eqref{magreshigh-N}. A natural extension of this work would be a detailed discussion of finite temperature away from charge neutrality in the whole parameter space. 

Our results are in a qualitative agreement with the main experimental findings on TMR, Ref.~\cite{RIS_1, PhysRevB.92.081306, RIS_5, 2016arXiv161001413C,PhysRevB.91.041203}, where a strong, linear TMR was observed
at finite carrier density in the ultra-quantum limit, which was comparable in magnitude to the Hall resistivity. 
The qualitative behavior of the TMR for shifted Weyl nodes (Fig.~\ref{fig:magnrescoul}) is similar to that observed in experiments where
pronounced Shubnikov - de Haas oscillations were superimposed on top of a rapidly growing background TMR. The order of magnitude of
the TMR in Fig.~\ref{fig:magnrescoul} is also comparable to the experimentally observed magnitude of the effect. 

Before concluding the paper, we briefly discuss alternative mechanisms of strong TMR that can emerge beyond the SCBA in the range
of magnetic field where magnetooscillations are strong. The first mechanism is based on classical memory effects (for reviews of memory effects 
in conventional 2D and 3D systems see Refs.~\cite{RevModPhys.84.1709} and \cite{MurzinUFN}, respectively). In conventional 3D systems 
a pronounced memory effect in a smooth disorder potential is based on the trapping of cyclotron orbits in $z$ direction \cite{polyakov}.
For the case of Weyl semimetals such a mechanism was recently addressed in Ref.~\cite{PhysRevB.92.180204}. This mechanism requires 
a large correlation radius of disorder $\xi$, which may be the case in two dimensions (large spacer) but seems unlikely in three dimensions, 
unless the ``fine structure constant'', Eq.~\eqref{fine} (assumed to be $\agt 1$ in the present work) is very small. Even within the assumption of $\xi$ being much larger than the cyclotron radius, Ref.~\cite{PhysRevB.92.180204} obtained the TMR up to 1-2 orders of magnitude, while it is of about 5 orders in the experiment Ref.~\cite{RIS_5}. Furthermore, in the ultra-quantum limit, this type of memory effects is expected to be strongly suppressed in Weyl systems, compared to conventional ones \cite{MurzinUFN}. This is due to the chirality of 1D modes in $z$ direction: the backscattering in $z$ direction requires internodal
scattering which is ineffective in Weyl materials (and also in Dirac semimetals in the strongest magnetic field which shifts the Dirac points in momentum space). An interesting prospect is to analyze quantitatively the role of this memory-effect mechanism of TMR in the case of screened Coulomb impurities in Weyl semimetals and compare it to the quantum TMR discussed in this
paper. Other mechanisms for a strong TMR can be provided by interaction effects (for this mechanism in 2D systems, see Ref. \cite{GM2004} and references therein), including possible Luttinger liquid effects of interaction within 1D channels in $z$ direction in the ultra-quantum limit, and by electron-hole recombination in compensated systems of a finite geometry (see Ref. \cite{Alekseev}). These mechanisms may be important in those regimes where the present model yields zero background TMR at moderate magnetic fields (higher Landau levels) and remain to be explored in realistic systems, in particular, in Weyl semimetals with shifted nodes. We do not expect, however, that these additional mechanisms of TMR would change the overall picture of TMR developed in the present work and could compete with quantum TMR in strongest magnetic fields.

\acknowledgments

We acknowledge interesting discussions with A. Andreev, J. Behrends, F. Kunst, J. Link, P. Ostrovsky, D. Polyakov, B. Sbierski, J. Schmalian, and B. Yan.
The work was supported by the EU Network FP7-
PEOPLE-2013-IRSES (project "InterNoM"), by Carl-Zeiss-Stiftung (J.K.), and by the Priority
Programme 1666 "Topological Insulators" of the Deutsche
Forschungsgemeinschaft (DFG-SPP 1666). 

\appendix
\begin{widetext}
\section{Anomalous Hall conductivity}
\label{app:anhall}
To evaluate the anomalous Hall conductivity, we calculate the particle density by integrating the density of states up to the ultraviolet cutoff:
\begin{align}
N(\mu, H)=\frac{\Omega^2}{8\pi^2v^3}\int_0^\Lambda d\varepsilon \left[f(\varepsilon-\mu)-f(\varepsilon+\mu)+1\right]\left( 1+2\sum_{n=1}^{\varepsilon^2/\Omega^2}\frac{\varepsilon}{\sqrt{\varepsilon^2-\Omega^2n}}\right),
\end{align}
where $f(\varepsilon\pm\mu)=(\exp((\varepsilon\pm\mu)/2T)+1)^{-1}$ denotes the Fermi function.
Taking the derivative with respect to $H$ leads to the following anomalous Hall conductivity:
\begin{align}
\sigma_{xy}^\text{II}=&\frac{e^2}{4\pi^2v}\left\lbrace\int_0^\infty d\varepsilon\left[f(\varepsilon-\mu)-f(\varepsilon+\mu)+1\right]\left( 1+2\sum_{n=1}^{\varepsilon^2/\Omega^2}\frac{\varepsilon}{\sqrt{\varepsilon^2-\Omega^2n}}\right)\right.
\notag
\\
&\left.+2\sum_{n=1}^{\Lambda^2/\Omega^2}\frac{\Omega^2n}{\sqrt{\Lambda^2-\Omega^2n}}-2\int_0^\infty d\varepsilon\left[\frac{df(\varepsilon-\mu)}{d\varepsilon}-\frac{df(\varepsilon+\mu)}{d\varepsilon}\right]
\sum_{n=1}^{\varepsilon^2/\Omega^2}\frac{\Omega^2n}{\sqrt{\varepsilon^2-\Omega^2n}}\right\rbrace.
\label{A2}
\end{align}
The last term of Eq.~(\ref{A2}) differs from the normal Hall conductivity Eq.~\eqref{normHallcalc} only by the sign.
The anomalous Hall conductivity reads
\begin{align}
\sigma_{xy}^\text{II}=\frac{e^2}{4\pi^2v}\left\lbrace \mu+2\sum_{n=1}^{\mu^2/\Omega^2}{\sqrt{\mu^2-\Omega^2n}}+\Lambda
+2\sum_{n=1}^{\Lambda^2/\Omega^2}{\sqrt{\Lambda^2-\Omega^2n}}
+2\sum_{n=1}^{\Lambda^2/\Omega^2}\frac{\Omega^2n}{\sqrt{\Lambda^2-\Omega^2n}}\right\rbrace-\sigma_{xy}^\text{I}.
\label{A3}
\end{align}
Employing the Euler Maclaurin formula to Eq.~(\ref{A3}) leads to a cancelation of the terms that depend on the ultraviolet cutoff $\Lambda$.
The remaining terms are
\begin{align}
\sigma_{xy}^\text{II}=\frac{e^2}{4\pi^2v}\left\lbrace \mu+2\sum_{n=1}^{\mu^2/\Omega^2}{\sqrt{\mu^2-\Omega^2n}}\right\rbrace-\sigma_{xy}^\text{I},
\end{align}
which corresponds to Eq.~\eqref{anhall}.

\section{Calculation of the normal Hall conductivity for large chemical potential}
\label{app:condhighmu}
This appendix is devoted to the calculation of the normal contribution to the Hall conductivity $\sigma_{xy}^{\text{I}}$
for large chemical potential, $\mu\gg\Omega$.
Starting from Eq.~\eqref{normHall}, we use the Green's functions for LLs with $n>0$. The resulting formula is
\begin{align}
\sigma_{xy}^{\text{I}}&=-\frac{2e^2\Omega^2}{(2\pi)^2v}\int d\varepsilon \frac{df(\varepsilon)}{d\varepsilon}\int\frac{dz}{2\pi}\sum_{n}\frac{\varepsilon\Gamma\Omega^2(z^2-\Gamma^2-\varepsilon^2)}{\left[\varepsilon^2-z^2-\Omega^2 n-\Gamma^2)^2+4\varepsilon^2\Gamma^2\right]\left[\varepsilon^2-z^2-\Omega^2 (n+1)-\Gamma^2)^2+4\varepsilon^2\Gamma^2\right]}
\end{align}
The evaluation of the integral over $z=vp_z$ leads to
\begin{align}
\sigma_{xy}^{\text{I}}&=-\frac{2e^2\Omega^2}{(2\pi)^2v}\int d\varepsilon \frac{df(\varepsilon)}{d\varepsilon}
\sum_{n}\text{Re}\left\lbrace\frac{1}{\Omega^2+4i\varepsilon\Gamma}
\left[\frac{-\Omega^2n-2\Gamma^2+2i\varepsilon\Gamma}
{\sqrt{\varepsilon^2-\Omega^2n-\Gamma^2+2i\varepsilon\Gamma}}+\frac{-\Omega^2(n+1)-2\Gamma^2+2i\varepsilon\Gamma}
{\sqrt{\varepsilon^2-\Omega^2(n+1)-\Gamma^2+2i\varepsilon\Gamma}}\right]\right\rbrace.
\end{align}

To simplify the equation, we can shift the sum over $n$ for the terms containing $n+1$ by $-1$ and evaluate the real part of the equation.
The Hall conductivity $\sigma_{xy}^{\text{I}}$ can be then written as
\begin{align}\label{app:Hall_calc}
\sigma_{xy}^{\text{I}}&=-\frac{2e^2\Omega^2}{(2\pi)^2v}\int d\varepsilon \frac{df(\varepsilon)}{d\varepsilon}
\left[\frac{-(2\Omega^2\Gamma^2-8\varepsilon^2\Gamma^2)}{\Omega^4(+4\varepsilon\Gamma)^2}\frac{\varepsilon-\Gamma}{\varepsilon^2+\Gamma^2}\right.
\nonumber\\
&\left.-\sum_{n=1}^{n_\text{max}-1}
\frac{\Omega^4n+2\Omega^2\Gamma^2}{\Omega^4+(4\varepsilon\Gamma)^2}
\frac{\sqrt{\varepsilon^2-\Omega^2n-\Gamma^2+\sqrt{(\varepsilon^2-\Omega^2-\Gamma^2)^2+4\varepsilon^2\Gamma^2}}}
{\sqrt{2}\sqrt{(\varepsilon^2-\Omega^2n-\Gamma^2)^2+4\varepsilon^2\Gamma^2}}
+\frac{4\varepsilon\Gamma\Omega\sqrt{n_\text{max}}}{\Omega^4+(4\varepsilon\Gamma)^2}\right].
\end{align}
We can split the sum over $n$ in three parts: $n<n_0$, $n_0$, $n_0+1$ and $n>n_0+1$, where $n_0$ is the resonant energy. For the part of $n<n_0$, we can neglect $\Gamma$ and for $n>n_0+1$ we can expand in $\Gamma$.
After some algebra, the Hall conductivity reads
\begin{align}\label{normalHallcalc}
\sigma_{xy}^{\text{I}}=&\frac{2e^2\Omega^2}{(2\pi)^2v}\int d\varepsilon \frac{df(\varepsilon)}{d\varepsilon}
\left[\frac{(2\Omega^2\Gamma^2-8\varepsilon^2\Gamma^2)}{\Omega^4(+4\varepsilon\Gamma)^2}\frac{\varepsilon-\Gamma}{\varepsilon^2+\Gamma^2}\right.
+\frac{2}{\Omega^4+(4\varepsilon\Gamma)^2}\sum_{n=1}^{n_0-1}
\frac{n}{\sqrt{\varepsilon^2-\Omega^2n}}
+\frac{2\Omega^4n_0}{\Omega^4+(4\varepsilon\Gamma)^2}
\frac{\Gamma^{(n_0)}}{A\varepsilon}
\nonumber
\\
&\left.
+\frac{2\Omega^4(n_0+1)}{\Omega^4+(4\varepsilon\Gamma)^2}
\frac{\Gamma^{(n_0+1)}}{A\varepsilon}
+\sum_{n=n_0+2}^{n_\text{max}-1}\frac{2\Omega^4n}{\Omega^4+(4\varepsilon\Gamma)^2}
\frac{2\varepsilon\Gamma}{\sqrt{\Omega^2n-\varepsilon^2}^3}
-\frac{4\varepsilon\Gamma\Omega\sqrt{n_\text{max}}}{\Omega^4+(4\varepsilon\Gamma)^2}\right].
\end{align}
Here $\Gamma^{(n_0)}$ and $\Gamma^{(n_0+1)}$ are defined via the self-consistent equation $\Gamma=\sum_n\Gamma^n$.
By evaluating the second sum, we see that term of the upper limit $n_\text{max}-1$ cancels with the last term of Eq.\eqref{normalHallcalc}.
Furthermore, the contribution of the lower limit $n_0+2$ of this sum and the term from the $n=0$ are parametrically small and can be neglected.
The normal contribution to the Hall conductivity then reads
\begin{align}
\sigma_{xy}^{\text{I}}=\frac{2e^2\Omega^2}{(2\pi)^2v}&\int d\varepsilon \frac{df(\varepsilon)}{d\varepsilon}\left[\frac{4}{3}\frac{\varepsilon^3}{\Omega^4+(4\varepsilon\Gamma)^2}
+\frac{\Omega^2\varepsilon^2}{\Omega^4+(4\varepsilon\Gamma)^2}\left(\frac{\Gamma}{A\varepsilon}-\frac{2\varepsilon}{\Omega^2}\right)\right].
\end{align}
This expression is further evaluated in the main text, where we consider the different regimes of LL broadening.
\end{widetext}

\section{Calculation of the magnetoresistance for pointlike impurities}\label{app:pointres}

In this appendix we evaluate the TMR for pointlike impurities.
For the lowest magnetic fields, $\Omega^2<\mu^3\gamma$, all LLs overlap. The conductivity and the normal Hall conductivity are given by the Drude formula in this regime, Eqs.~\eqref{conover} and \eqref{Hallover}, leading to a vanishing TMR. In this regime, the anomalous Hall conductivity is exponentially small, see Eq.~\eqref{Nover}. Therefore, effects of a finite temperature (not discussed here) will dominate the TMR.
For magnetic fields in the range $\mu^3\gamma<\Omega^2<\mu^{5/2}\gamma^{1/2}$, the LLs are separated, but the background density of states is still larger than the peaks of the LLs. In this region, the conductivity, Eq.~\eqref{conback}, is smaller than the Hall conductivity, Eq.~\eqref{Hall}. The magnetoresistance calculated with Eq.~\eqref{rho-by-xy} remains zero.

A further increase of magnetic field, $\mu^{5/2}\gamma^{1/2}<\Omega^2<\mu^2$, leads to pronounced LLs. The TMR is still determined by Eq.~\eqref{rho-by-xy} (conductivity is small compared to the Hall conductivity), but it now strongly oscillates with magnetic field because of the oscillations of the scattering rate. With the conductivity, Eq.~\eqref{consep}, and the Hall conductivity, Eq.~\eqref{Hall}, the TMR is evaluated as
\begin{align}\label{rhosep}
\Delta_\rho(H)\sim\frac{\Gamma^2(\mu)}{\gamma^2\mu^4}-1
\end{align}
leading to 
\begin{align}\label{rhopeak}
\Delta_\rho^\text{max}(H)\sim\frac{\Omega^{8/3}}{\mu^{10/3}\gamma^{2/3}}
\end{align}
at the peak (using the conductivity at the peak, Eq.~\eqref{conpeak})
and zero background TMR (as in the previous region).

For stronger magnetic fields, $\mu<\Omega<\gamma^{-1}$, the TMR is determined by carriers at the zeroth LL. With the conductivity, Eq.~\eqref{con1}, and the Hall conductivity, Eq.~\eqref{Hall0}, we find that $\sigma_{xy}$ is larger than $\sigma_{xx}$ up to magnetic fields of $\Omega^2<\mu\gamma^{-1}$, resulting in
\begin{align}\label{rho0small}
\Delta_\rho\sim\frac{\Omega^2}{\mu^2}-1,
\end{align}
where we have used Eq.~\eqref{rho-by-xy}. 
For yet higher magnetic fields, $\mu^{1/2}\gamma^{-1/2}<\Omega<\gamma^{-1}$, we use Eq.~\eqref{rho-by-xx} 
and obtain
\begin{align}\label{rho0large}
\Delta_\rho\sim\frac{1}{\gamma^2\Omega^2}-1.
\end{align}

We continue this appendix with the analysis of the TMR for a fixed particle density. 
The particle density is evaluated with Eq.~\eqref{partdens}, reading
\begin{align}\label{partdens-mu}
N(\mu,\Omega)=\left\lbrace\begin{array}{ll}
\dfrac{\mu\Omega^2}{4\pi^2 v^3} & \qquad \Omega>\mu,\\[0.2cm]
\dfrac{\mu^3}{12\pi^2 v^3} & \qquad \Omega<\mu.
\end{array}\right.
\end{align}
The magnetic-field dependence of the resistivity only changes for the zeroth LL (in both conductivity and Hall conductivity). 
For completeness, we start with the analysis from the lowest relevant magnetic fields, $N^{5/6}\gamma^{1/2}>\Omega^2$ (below no TMR emerges to the leading order within the SCBA).

In magnetic fields $N^{5/6}\gamma^{1/2}<\Omega^2<N^{2/3}$ the TMR is finite at the center of the LLs, Eq.~\eqref{rhopeak}. 
Using Eq.~\eqref{partdens-mu}, we get 
\begin{align}\label{rhopeak-N}
\Delta_\rho^\text{max}(H)\sim\frac{\Omega^{8/3}}{N^{10/9}\gamma^{2/3}}
\end{align}
for the TMR at the center of LLs for a fixed particle density

For larger magnetic fields, $N^{1/3}<\Omega<\gamma^{-1}$, the conductivity, Eq.~\eqref{con1}, and the Hall conductivity, Eq.~\eqref{Hall0}, are modified by Eq.~\eqref{partdens-mu}. We find that the Hall conductivity is larger than the conductivity up to magnetic fields of $\Omega<N^{1/4}\gamma^{-1/4}$, resulting in \begin{align}
\Delta_\rho\sim\frac{\Omega^6}{N^2}-1.
\end{align}
For yet stronger magnetic fields, $N^{1/4}\gamma^{-1/4}<\Omega<\gamma^{-1}$, we use Eq.~\eqref{rho0large} which remains unaffected for a fixed particle density.

The calculation of the magnetoresistance was so far limited to zero temperature. In the following, we will briefly discuss the effect of finite temperatures. Finite temperature smears LLs for $T>\Omega/\sqrt{n}$. Let us consider separated LLs in the regime of low chemical potential, $\Omega<\mu<\Omega(\Omega/A)^{1/5}$, and temperature $T<\mu$. 
In this case, the contribution of the LLs in the vicinity of the 
chemical potential $\mu-T<\Omega \sqrt{n}<\mu+T$ should be analyzed. In order to estimate the corresponding contribution to the conductivity,
we replace the integral over energy by a sum over regions of width $\Gamma(W_n)$ 
around Landau levels, and replace $\Gamma^{(n)}(\epsilon)$ there by its maximal value 
$\Gamma^{(n)}(W_n)\equiv \Gamma_n \sim A^{2/3}\Omega^{1/3} n^{1/6}$.
As a result, we get
\begin{eqnarray}
\sigma_{xx}^{(n)}
&\sim& \frac{e^2\Omega^2}{A T v} \sum_{n=\mu(\mu-T)/\Omega^2}^{\mu(\mu+T)/\Omega^2}
\Gamma_n\ \frac{\Gamma_n^2 W_n^2}{(4 W_n \Gamma_n)^2+\Omega^4}\notag
\\
& \sim &\frac{e^2 \gamma \mu^4}{\Omega^2 v^4}
\propto \frac{ \gamma \mu^4}{H}.
\end{eqnarray}
This value of the conductivity is smaller than the background conductivity Eq.~\eqref{conback}, but is important for the TMR which otherwise vanishes.  

The Hall conductivity for $T<\mu$ remains essentially unaffected by finite temperature. The magnetoresistance is still determined by the Hall conductivity according to Eq.~\eqref{rho-by-xy}, yielding Eq.~\eqref{rho_finiteT}. This linear magnetoresistance is small and will show exponentially suppressed Shubnikov-de Haas oscillations.

\section{Calculation of the conductivity and magnetoresistance for charged impurities}\label{App:charged-xx-xy}

In this appendix we present details of evaluation of the TMR for Coulomb impurities.
The disorder strength $\gamma$ is substituted for charged impurities by
\begin{align}
\gamma(\mu,H)=\left\lbrace\begin{array}{cc}
\dfrac{\varepsilon^3_\text{imp}}{\mu^4}, &\mu>\Omega,\varepsilon_\text{imp}\\[0.4cm]
\dfrac{\varepsilon^3_\text{imp}}{\Omega^4},& \Omega>\mu, \varepsilon_\text{imp}\\[0.4cm]
\dfrac{1}{\varepsilon_\text{imp}}, & \varepsilon_\text{imp}>\mu,\Omega
\end{array}\right.
\end{align}
As discussed in the main text, this substitution does not provide exact numerical prefactors which are  neglected in the following.
The Hall conductivity remains unaffected by this transformation since it does not depend on disorder.
The most important transformation is performed for the lowest LL, $\Omega>\mu$, where the screening depends on magnetic field. With $\gamma\to\varepsilon_\text{imp}^3\Omega^{-4}$, the conductivity for the zeroth LL, Eq.~\eqref{con1}, transforms to
\begin{align}\label{concoul1}
\sigma_{xx}\sim\frac{e^2\varepsilon_\text{imp}^3}{v\Omega^2}.
\end{align}
For higher LLs, $\mu>\Omega$, the self-consistent equation for the LLs broadening, Eq.~\ref{GammanAPP}, is transformed into
\begin{align}
&\Gamma_C(\mu,\Omega)\sim\frac{\varepsilon_\text{imp}^3\Omega^2}{\mu^3}\nonumber \\
&\quad \times\sum_n\frac{\sqrt{\mu^2-W_n^2+\sqrt{(W_n^2-\mu^2)^2+4 \mu^2 \Gamma^2_C}}}{\sqrt{2}\  \sqrt{(W_n^2-\mu^2)^2+4 \mu^2 \Gamma^2_C} }.
\nonumber
\\
\label{SCBAcoul}
\end{align}
The full equation is solved similarly to the case of pointlike impurities. 
Most interesting are the value of $\Gamma_C$ at the center of a particular LL and the background value of $\Gamma_C$:
\begin{align}
\Gamma_C^\text{peak}\sim \frac{\varepsilon^2_\text{imp}\Omega^{4/3}}{\mu^{8/3}}\varepsilon^{1/3},
\qquad
\Gamma_C^\text{bg}\sim \frac{\varepsilon^3_\text{imp}}{\mu^3}\varepsilon.
\end{align}

The conductivity of separated LLs is given by Eq.~\eqref{consep} which is valid for $\mu^{3/2}\gamma^{1/2}<\Omega<\mu$.
For charged impurities, the conductivity transforms as follows
\begin{align}\label{concoulsep}
\sigma_{xx}\sim \frac{\mu^2\Gamma^2}{A\Omega^2}\qquad\rightarrow\qquad\sigma_{xx}\sim\frac{\mu^6\Gamma_C^2}{\varepsilon_\text{imp}^3\Omega^4}.
\end{align}
The background and peak values of the conductivity then read:
\begin{align}\label{concoulpeak}
\sigma_{xx}^\text{bg}\sim \frac{\mu^2\varepsilon^3_\text{imp}}{\Omega^4},
\qquad\sigma_{xx}^\text{peak}\sim\frac{\mu^{4/3}\varepsilon_\text{imp}}{\Omega^{4/3}}.
\end{align}
The background contribution is important in the whole range of separated LLs, while the peaks are only present for $\mu^{5/4}\gamma^{1/4}<\Omega<\mu$.

The conductivity for overlapping LLs is given, in the case of pointlike impurities, by Eq.~\eqref{conover}. 
For the transformation to Coulomb impurities, we need to distinguish between the range of high chemical potential $\mu\gg \varepsilon_\text{imp}$, where
\begin{align}
\sigma_{xx}\sim \frac{e^2\mu^4}{v\varepsilon_\text{imp}^3},
\end{align}
and the range of lower chemical potential, $\mu<\varepsilon_\text{imp}$, where
\begin{align}
\sigma_{xx}\sim \frac{e^2\varepsilon_\text{imp}}{v}.
\end{align}

The results for the conductivity in the case of charged impurities are summarized as follows: 
\begin{align}\label{sigma-xx-coul-low}
\sigma_{xx}\sim \frac{e^2}{v}\left\{
  \begin{array}{ll}
  \varepsilon_\text{imp},  &\quad \Omega\ll\varepsilon_\text{imp},\\[0.4cm]
    \dfrac{\varepsilon_\text{imp}^3}{\Omega^2}, &\quad \varepsilon_\text{imp}\ll\Omega,
    \end{array}\right.
\end{align}
for $\mu\ll \varepsilon_\text{imp}$ and
\begin{align}\label{sigma-xx-coul-high}
\sigma_{xx}\sim \frac{e^2}{v}\left\{
  \begin{array}{ll}
    \dfrac{\mu^4}{\varepsilon_\text{imp}^3},  &\quad \Omega\ll\dfrac{\varepsilon_\text{imp}^{3/2}}{\mu^{1/2}},\\[0.4cm]
   \dfrac{\mu^6\Gamma_\text{C}^2(\mu)}{\varepsilon_\text{imp}^3\Omega^4},  &\quad \dfrac{\varepsilon_\text{imp}^{3/2}}{\mu^{1/2}}\ll\Omega<\mu,\\[0.4cm]
    \dfrac{\varepsilon_\text{imp}^3}{\Omega^2}, &\quad \mu<\Omega
    \end{array}\right.
\end{align}
for $\mu\gg\varepsilon_\text{imp}$.

For the evaluation of the TMR, we compare $\sigma_{xx}$ and $\sigma_{xy}$.
We start with small chemical potential, $\mu<\varepsilon_\text{imp}$. 
For $\Omega\ll\varepsilon_\text{imp}$, the LLs overlap, leading to an exponentially small TMR, as discussed in Appendix~\ref{app:pointres}. 
For $\Omega\gg\varepsilon_\text{imp}$, the conductivity, Eq.~\eqref{concoul1}, is larger than the Hall conductivity Eq.~\eqref{Hall0}, for $\Omega\ll\varepsilon_\text{imp}^{3/2}\mu^{-1/2}$, leading to 
\begin{align}
\Delta_\rho(H)\sim \frac{\Omega^2}{\varepsilon_\text{imp}^2}-1.
\end{align}
For largest magnetic fields, $\Omega\gg\varepsilon_\text{imp}^{3/2}\mu^{-1/2}$, the Hall conductivity dominates, resulting in
\begin{align}
\Delta_\rho(H)\sim \frac{\varepsilon_\text{imp}^4}{\Omega^2\mu^2}-1
\end{align}
for the TMR.

In the limit $\mu\gg\varepsilon_\text{imp}$, a finite TMR is found for fields larger than $\mu^{1/4}\varepsilon_\text{imp}^{3/4}$. 
For lower fields, either the LLs are overlapping or the background of separated LLs dominates the density of states. 
For both cases, the TMR is negligibly small for the reasons discussed in Appendix~\ref{app:pointres}.
For $\mu^{1/4}\varepsilon_\text{imp}^{3/4}\ll\Omega\ll\mu$, the Hall conductivity, Eq.~\ref{Hall}, is still larger than the background $\sigma_{xx}$, 
but at the center of the LLs we find a finite TMR corresponding to the peak value of the conductivity, Eq.~\eqref{concoulpeak}:
\begin{align}
\Delta_\rho(H)\sim\frac{\Omega^{8/3}}{\mu^{2/3}\varepsilon_\text{imp}^2}-1.
\end{align}

In largest magnetic fields, $\Omega>\mu$, the TMR of the zeroth LLs is determined by the Hall conductivity, because $\sigma_{xy}>\sigma_{xx}$ for $\Omega>\varepsilon_\text{imp}^{3/2}\mu^{-1/2}$ [cf. Eq.~\eqref{concoul1} and \eqref{Hall0}] which is fulfilled in the whole regime of the zeroth LL. 
Therefore, the TMR is given by (using Eq.~\eqref{rho-by-xy})
\begin{align}
\Delta_\rho(H)\sim\frac{\mu^2}{\Omega^2}-1.
\end{align}

We continue this appendix with the discussion of a fixed particle density. 
The particle density remains unaffected by charged impurities and is given by Eq.~\eqref{partdens-mu}. 
For $N^{1/3}\ll \varepsilon_\text{imp}$ and $\Omega\gg \varepsilon_\text{imp}$, the conductivity of the zeroth LL, Eq.~\eqref{concoul1}, is larger than $\sigma_{xy}$, resulting in
\begin{align}
\Delta_\rho(H)\sim \frac{\Omega^2}{\varepsilon_\text{imp}^2}-1.
\end{align}
In the limit $N^{1/3}\gg\varepsilon_\text{imp}$, a finite TMR is found for fields larger then $N^{1/12}\varepsilon_\text{imp}^{3/4}$:
\begin{align}
\Delta_\rho(H)\sim\frac{\Omega^{8/3}}{N^{2/9}\varepsilon_\text{imp}^2}-1.
\end{align}

In strongest magnetic fields $\Omega>N^{1/3}$, the comparison of the conductivity Eq.~\eqref{concoul1} and the Hall conductivity, Eq.~\eqref{Hall0}, shows that the Hall conductivity  dominates in the whole regime ($\Omega>\varepsilon_\text{imp}^{3/2}N^{-1/6}$). Therefore, the TMR is given by (using Eq.~\eqref{rho-by-xy})
\begin{align}
\Delta_\rho(H)\sim\frac{\Omega^2}{N^{2/3}}-1.
\end{align}
These results are summarized in Fig. \ref{phasediagramCoul} of the main text.

\end{document}